\documentclass[a4paper,fleqn,usenatbib]{mnras}

\pdfoutput=1

\usepackage{graphicx}	
\usepackage{amsmath}	
\usepackage{amssymb}	

\usepackage[T5]{fontenc} 
\usepackage{aecompl}
\def\pasj{PASJ}

\title[GW Librae: The laboratory for pulsations]{GW Librae: A unique laboratory for pulsations in an accreting white dwarf}

\author[Toloza et al.]{O. ~Toloza,$^{1}$\thanks{O.F.C.Toloza@warwick.ac.uk}
B.~T.~G\"{a}nsicke,$^{1}$
J.~J.~Hermes,$^{2,19}$
D.~M.~Townsley,$^{3}$
\newauthor
M.~R.~Schreiber$^{4}$,
P.~Szkody$^{5}$,
A.~Pala$^{1}$,
K.~Beuermann$^{6}$,
L.~Bildsten$^{7}$,
\newauthor
E.~Breedt$^{1}$,
M.~Cook$^{8,10}$,
P.~Godon$^{9}$,
A.~A. Henden$^{10}$,
I.~Hubeny$^{11}$,
\newauthor
C.~Knigge$^{12}$,
K.~S.~Long$^{13}$,
T.~R.~Marsh$^{1}$,
D.~de Martino$^{14}$,
\newauthor
A.~S.~Mukadam$^{5,15}$,
G.~Myers$^{10}$,
P.~Nelson $^{16}$,
A.~Oksanen$^{9, 17}$,
\newauthor
J.~Patterson$^{18}$,
E.~M.~Sion$^{9}$,
M.~Zorotovic$^{4}$
\\
$^{1}$Department of Physics, University of Warwick, Coventry\,-\,CV4~7AL, UK\\
$^{2}$Department of Physics and Astronomy, University of North Carolina, Chapel Hill, NC 27599-3255, USA\\
$^{3}$Department of Physics and Astronomy, The University of Alabama, Tuscaloosa, AL\,-\,35487, USA\\
$^{4}$Instituto de F\'isica y Astronom\'ia, Universidad de Valpara\'iso, Valpara\'iso, 2360102, Chile\\
$^{5}$Department of Astronomy, University of Washington, Seattle, WA\,-\,98195-1580, USA\\
$^{6}$Institut f\"ur Astrophysik Friedrich-Hund-Platz 1, Georg-August-Universit\"at, G\"ottingen, 37077, Germany\\
$^{7}$Kavli Institute for Theoretical Physics and Department of Physics Kohn Hall, University of California, Santa Barbara, CA~93106, USA\\
$^{8}$Newcastle Observatory, Newcastle, ON L1B 1M5 Canada\\
$^{9}$Astrophysics and Planetary Science, Villanova University, Villanova, PA~19085, USA\\
$^{10}$American Association Of Variable Star Observers\\
$^{11}$Steward Observatory and Dept. of Astronomy, University of Arizona, Tucson, AZ~85721, USA\\
$^{12}$School of Physics and Astronomy, University of Southampton, Southampton,	SO17~1BJ, UK\\
$^{13}$Space Telescope Science Institute, Baltimore, MD 21218, USA\\
$^{14}$Osservatorio Astronomico di Capodimonte, salita Moiariello 16, 80131 Napoli, Italy\\
$^{15}$Apache Point Observatory, 2001 Apache Point Road, Sunspot, NM 88349-0059, USA\\
$^{16}$1105 Hazeldean Rd, Ellinbank 3820, Australia\\
$^{17}$Caisey Harlingten Observatory, San Pedro de Atacama, Chile\\
$^{18}$Department of Astronomy, Columbia University, New York, NY~10027, USA\\
$^{19}$Hubble Fellow\\
}

\date{Accepted XXX. Received YYY; in original form ZZZ}

\pubyear{2016}

\begin{document}
\label{firstpage}
\pagerange{\pageref{firstpage}--\pageref{lastpage}}
\maketitle

\begin{abstract}
Non-radial pulsations have been identified in a number of accreting white dwarfs in cataclysmic variables. These stars offer insight into the excitation of pulsation modes in atmospheres with mixed compositions of hydrogen, helium, and metals, and the response of these modes to changes in the white dwarf temperature. Among all pulsating cataclysmic variable white dwarfs, GW\,Librae stands out by having a well-established observational record of three independent pulsation modes that disappeared when the white dwarf temperature rose dramatically following its 2007 accretion outburst. Our analysis of {\em HST} ultraviolet spectroscopy taken in 2002, 2010 and 2011, showed that pulsations produce variations in the white dwarf effective temperature as predicted by theory. Additionally in May~2013, we obtained new {\em HST}/COS ultraviolet observations that displayed unexpected behaviour: besides showing variability at $\simeq$275\,s, which is close to the post-outburst pulsations detected with {\em HST} in 2010 and 2011, the white dwarf exhibits high-amplitude variability on a $\simeq$4.4\,h time-scale. We demonstrate that this variability is produced by an increase of the temperature of a region on white dwarf covering up to $\simeq$30 per cent of the visible white dwarf surface. We argue against a short-lived accretion episode as the explanation of such heating, and discuss this event in the context of non-radial pulsations on a rapidly rotating star.
\end{abstract}

\begin{keywords}
binaries: close -- Stars: white dwarfs -- variables: general -- stars: individual (GW Librae)
\end{keywords}



\section{Introduction}

Cataclysmic variables (CVs) are interacting binaries composed of a white dwarf accreting from a close, non-degenerate companion \citep{warner95-1}. Dwarf novae (DNe) are a subset of CVs that have low secular accretion rates and thermally unstable accretion discs. These instabilities occasionally trigger outburst events during which the systems substantially brighten \citep{meyer+meyer-hofmeister81-1}. However, DNe are in quiescence most of the time, and for short period systems in this state of low mass transfer the white dwarf often dominates their ultraviolet and sometimes their optical spectra \citep[e.g.][]{szkody+mattei84-1,sionetal94-1,gaensickeetal05-2}.

A handful of quiescent DNe show photometric variability at periods in the range 100$-$1900\,s, consistent with non-radial $g$-mode pulsations on the white dwarf \citep{Mukadametal06-1}. These accreting pulsating white dwarfs have hydrogen-dominated atmospheres, but are generally too hot to fall within the instability strip of isolated white dwarfs \citep{Szkodyetal10-2}, which empirically extends from $\simeq$11\,100$-$12\,600\,K for ZZ~Ceti stars \citep{gianninasetal11-1}. However, ongoing accretion from a low-mass companion may result in an enhanced helium abundance, which can drive oscillations at higher temperatures, 15\,000$-$20\,000\,K, as a result of a subsurface He\,{\sc ii} partial ionization zone \citep{arrasetal06-1, VanGrooteletal2015}.

Non-radial pulsations in accreting white dwarfs are potentially useful to learn about their internal structure using asteroseismic techniques. In particular, CV white dwarfs are heated during outbursts \citep[e.g.][]{Godonetal2003-1, piroetal05-1}, and subsequently cool back to their quiescent temperature. As a result, accreting white dwarf pulsators can evolve through the instability strip in a few years, \textit{much} faster than their isolated counterparts, which cool through the strip on evolutionary time scales of 5$-$10$\times$10$^{8}$ yr. Just as with isolated white dwarfs, we expect that as a pulsating white dwarf cools, its convection zone deepens, driving longer-period pulsations \citep{Brickhill91-2,Mukadametal06-1}.

GW\,Librae (GW\,Lib) is the prototypical DN that exhibits non-radial pulsations in quiescence \citep{vanzyletal00-1}. These authors reported optical pulsation periods near 650, 370, and 230\,s that, with the exception of the 230\,s mode, remained fairly constant for years after discovery \citep{vanzyletal04-1}. Follow-up {\em Hubble Space Telescope} ({\em HST}) ultraviolet observations confirmed the pulsation origin of this optical variability: the Fourier transform of the ultraviolet light curve shows periodic signals at 646, 376, and 237\,s, but with amplitudes roughly 10 times higher than in the optical \citep{szkodyetal02-4}. Such a wavelength-dependent amplitude is expected because the ultraviolet lies in the exponential Wien tail of the spectral energy distribution, where the flux is more sensitive to changes in the temperature  than at optical wavelengths \citep{Robinsonetal95-2}. Additionally the amplitude depends also on the wavelength-dependent limb darkening. Towards optical wavelengths the limb darkening decreases, increasing the geometric cancellation effects, therefore the amplitudes are highly reduced \citep{Robinsonetal95-2}.

GW\,Lib underwent a large-amplitude outburst in 2007~April which took its quiescent magnitude from $V\simeq$17\,mag \citep{thorstensenetal02-2} to $V\simeq$8\,mag \citep{Templetonetal2007-cbet}. The stable pulsation signals reported by \citet{vanzyletal00-1} were immediately swamped by the light from the accretion disk and an orbital superhump near 80\,min became the dominant variability in the optical light curve \citep{katoetal2008, Bullocketal11-1}. To date, several prominent signals have recurrently shown up \citep{copperwheatetal09-1, schwieterman2010, Bullocketal11-1, Vicanetal11-1, szkodyetal12-1}. Except for a signal near 290$-$300\,s, all other signals have relatively long periods ($\sim$19\,min, $\sim$ 2\,h, and $\sim$4\,h).

A 296\,s periodicity was first detected on 21th Jun~2008 in optical photometry obtained with the high speed ULTRACAM photometer \citep{copperwheatetal09-1}. {\em HST} ultraviolet observations in 2010 revealed multiple closely spaced periodicities, but with the highest peak at 292\,s. The signal was also detected in 2011 with {\em HST} near 293\,s, with an amplitude more than twice as large as in 2010 \citep{szkodyetal12-1}. If this signal is the return of the $\sim$237\,s period after outburst, it demonstrates the unique opportunity that DN outbursts offer to study the evolution of the pulsation spectrum as a function of white dwarf cooling. 

The $\simeq$19\,min signal was first detected from mid-March to July in 2008 in the optical \citep{copperwheatetal09-1, schwieterman2010, Vicanetal11-1, Bullocketal11-1}, returning in May~2012 \citep{choteetal-2016}. Although its origin was associated with an instability in the accretion disc \citep{Vicanetal11-1} another possibility is that it could be a different pulsation mode driven by the white dwarf \citep{choteetal-2016}.

The $\simeq$2\,h signal has been detected prior- and post-outburst. It was first seen in May~2001 in the optical by \citet{woudt+warner02-2}. In May~2005, it was identified in optical data by \citet{copperwheatetal09-1} and again in August by \citet{Hilton2007}. Also it was reported in post outburst optical data in 2008 \citep{Vicanetal11-1}, confirming it to be a recurrent feature, though not always present in photometric observations.

The $\simeq$4\,h variability has been observed at optical and ultraviolet wavelengths \citep{schwieterman2010, Bullocketal11-1, Vicanetal11-1}. \citet{Bullocketal11-1} obtained ultraviolet (GALEX FUV and NUV) and ground-based optical photometry over a period of three years following the 2007 outburst. In their ultraviolet observations, a $\simeq$4\,h variability made its first apparition in 2008, increasing in amplitude during the following two years. In the optical, the variability was intermittently detected in 2009. \citet{Vicanetal11-1} performed an intensive ground based campaign of time series photometry covering the 2007 outburst, as well as two different years after the outburst, 2008 and 2010. They showed that, in fact, the $\simeq$4\,h variability was present in 2008, and became stronger in 2010. In addition, \citet{schwieterman2010} also reported a $\simeq$4\,h signal detected in their photometric data taken in 2008. The three groups agree that this is a recurrent signal that wanders in phase and amplitude on time scales of days and occasionally disappears. These detailed photometric studies presented in those papers suggest that the detected $\simeq$4\,h periodicity is, in fact, the fundamental of the $\simeq$2\,h signal, that was repeatedly detected earlier.

In this paper, we present new observations of GW\,Lib performed with the Cosmic Origin Spectrograph (COS) onboard {\em HST} on 2013 May 30. In Section \ref{observations}, we describe the observations and the light curve analysis, revealing the presence of a large amplitude variability with period similar to the $\simeq$4\,h signal detected previously, which we discussed above. In Section \ref{method}, we outline the spectral fitting procedure we performed to determine atmospheric parameters, emphasizing the use of the Markov Chain Monte Carlo method. We discuss the results in Section \ref{discussion}, and summarise our conclusions in Section \ref{conclusions}

\section{OBSERVATIONS}
\label{observations}

\begin{table}  
\begin{center}
\caption[Observations]{Log of all {\em HST} observations of GW\,Lib taken in time-tag mode and using the G140L gratings. The central wavelength was set to 1425\,\AA\ and 1105\,\AA\ for the STIS and COS observations, respectively. \label{tab:obs}}
\begin{tabular}{c c c c c}
\hline
\hline
 instrument & orbit & date	& UT Time 	& Exp. Time (s)	\\
\hline
STIS	&1& 2002-01-17 & 01:52:00 & 2105\\
STIS	&2& 2002-01-17 & 03:14:04 & 2603\\
STIS	&3& 2002-01-17 & 04:50:15 & 2603\\
STIS	&4& 2002-01-17 & 06:26:25 & 2580\\
\hline
COS & 5 & 2010-03-11 & 10:16:45 & 2906  \\
\hline
COS & 1 & 2011-04-09 & 14:07:50 & 2128  \\
COS & 2 & 2011-04-09 & 15:43:41 & 2971  \\
\hline
COS	&1 & 2013-05-30	&11:41:59	&2182	\\
COS	&2 & 2013-05-30	&13:11:34	&1746	\\
COS	&2 & 2013-05-30	&13:42:25	&876	\\
COS	&3 & 2013-05-30	&14:47:15	&864	\\
COS	&3 & 2013-05-30	&15:03:34	&1748	\\
\hline
\end{tabular}
\end{center}
\end{table}

\subsection{Ultraviolet spectroscopy}

\begin{figure}
\centering{\includegraphics[width=\columnwidth]{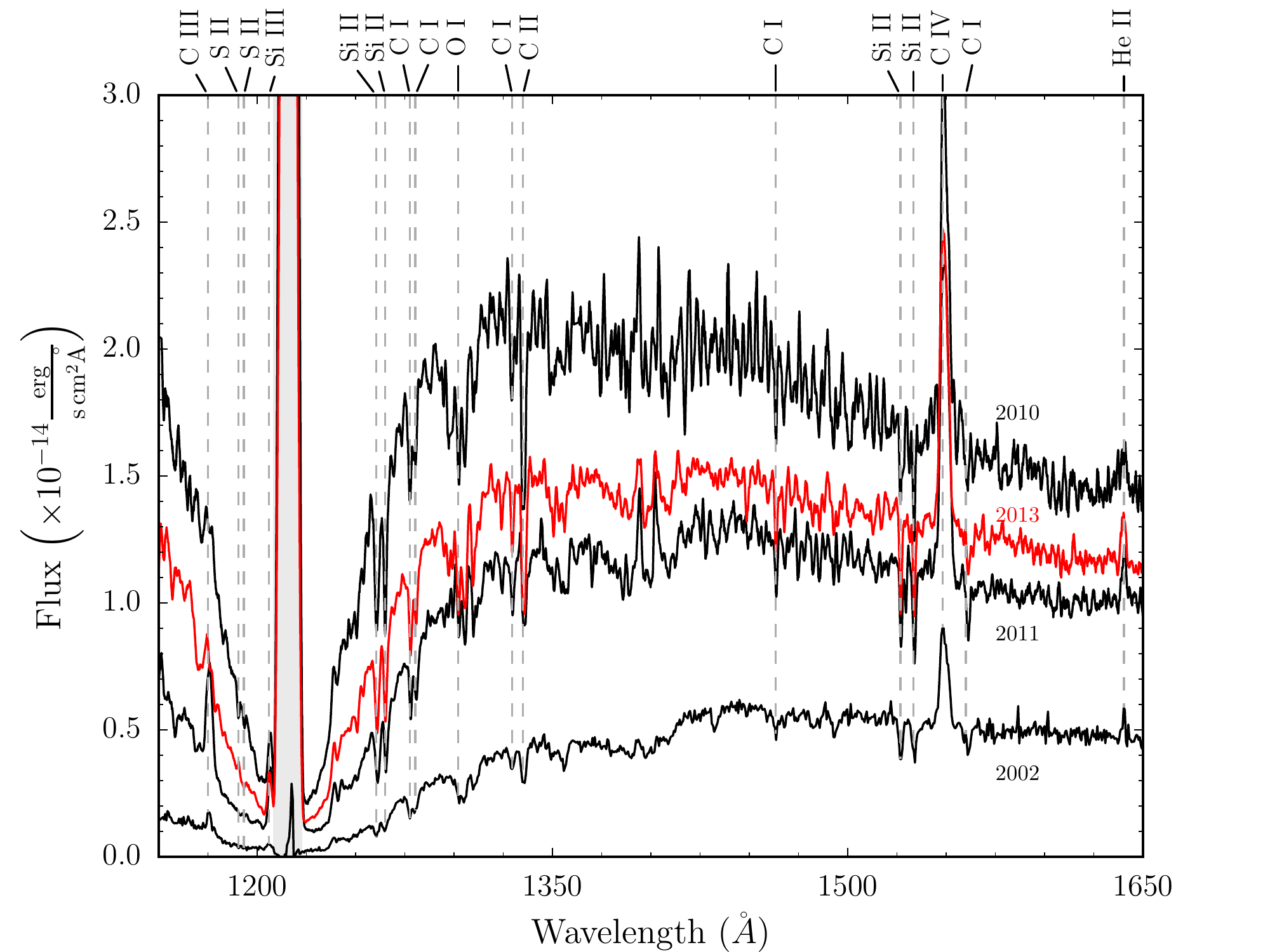}}
\caption{Average COS/G140L and STIS/G140L spectra of GW\,Lib (no flux offsets were applied). Black lines from top to bottom show the observations from 2010, 2011 (both COS), and 2002 (STIS). The red line corresponds to the new 2013 COS observations, which unexpectedly show a higher flux level than the 2011 observations. The most prominent lines are labelled and the airglow emission Ly$\alpha$ at 1216\,\AA\ is shaded in light grey. \label{all_spectra}}
\end{figure}

We observed GW\,Lib as part of a far-ultraviolet COS survey to determine the mass and temperature distribution of accreting white dwarfs (Cycle 20 programme 12870). We obtained  a total of 123.6\,min of time-tagged spectroscopy of GW\,Lib on 2013~May 30 over three consecutive orbits using the G140L gratings, which covers the wavelength range $1150-1800$\,\AA\ with roughly $0.75$\,\AA\ resolution (red line in Fig.\,\ref{all_spectra}).

Additionally, we retrieved from the {\em MAST} archive all Space Telescope Imaging Spectrograph (STIS) and COS ultraviolet data taken with the G140L grating, and using the time-tag mode: four STIS orbits taken prior to the 2007 outburst (for more details see \citealt{szkodyetal02-4}), and a total of three COS orbits post-outburst data (see \citealt{szkodyetal12-1}). Table \ref{tab:obs} summarizes the observations and the time-averaged spectrum of each observation is shown in Fig.\,\ref{all_spectra}. 

The time-averaged spectrum of each epoch is dominated by the white dwarf. However, it is clear that there is some flux contribution from an additional continuum component since the core of Ly$\alpha$ does not reach zero. This second component seems to be flat and without significant features. Such a flux component has been identified in {\em HST} observations of many other DNe (e.g. VW\,Hyi, \citealt{godonetal2004-2, longetal2009-1}; VY\,Aqr \& WX\,Ceti, \citealt{sionetal03-1}; SW UMa \& BC UMa, \citealt{gaensickeetal05-2}). In the case of VW\,Hyi a flat and featureless continuum contributing at shorter wavelengths has also been detected with FUSE ($<$ 970.8\,\AA; \citealt{godon2008A-1, longetal2009-1}). In all cases the flux contribution of this second component is small ($\lesssim$ 20 per cent). The exact origin of that second component is not well understood; possible locations are the hot innermost region of the accretion disc, or a boundary/spreading layer on the white dwarf \citep{godonetal95, godon2008A-1}. 

Following the 2007 outburst, the white dwarf in GW\,Lib is expected to gradually cool while it is relaxing to its quiescent state. However, the average flux observed in 2013 is higher than in 2011. 

\subsection{Variability}

\begin{figure*}
\centering{\includegraphics[width=2\columnwidth]{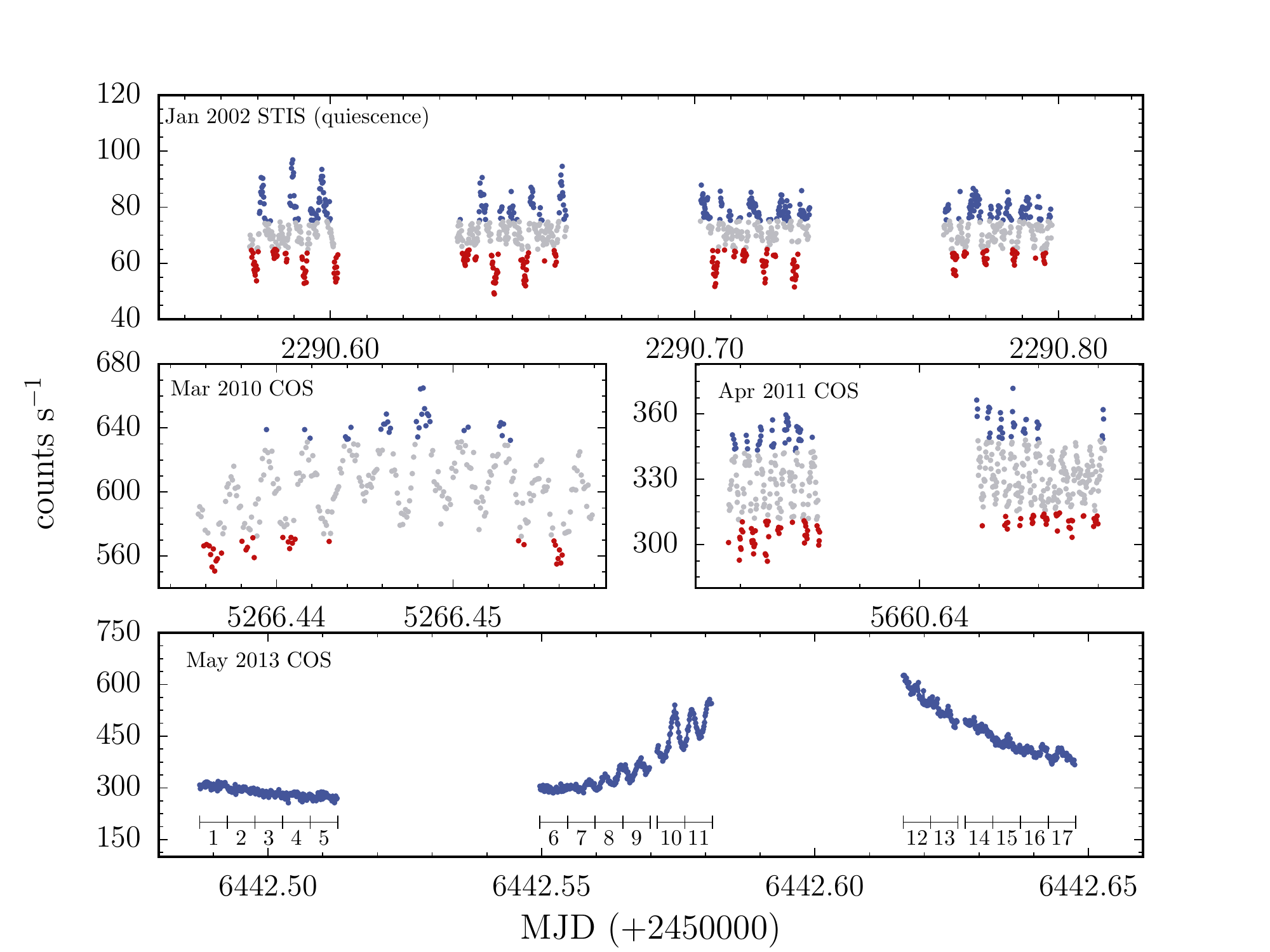}}
\caption{Light curves of GW\,Lib constructed from the time-tag data in 10\,s bins. The upper panel shows the STIS/G140L pre-2007 outburst light curve while the other panels correspond to COS/G140L post-outburst observations. The count rate strongly fluctuates in the 2013 observations, increasing likely beyond 600 counts\,s$^{-1}$ during the gap in the data (comparable to the mean level in 2010). In addition, short-term oscillations are clearly visible during the second orbit. From the observations obtained in 2002, 2010, and 2011, we extracted two spectra, one \textit{peak spectrum} constructed from the intervals with the highest count rate (blue dots), and one \textit{trough spectrum} from those with the lowest count rate (red dots).  For the 2013 observation a set of 17 spectra was created, as labelled underneath the light curve. We performed spectral fits to these spectra to investigate variations in the effective temperature (see Section \ref{method} for details).\label{lc}}
\end{figure*}

We are able to construct ultraviolet light curves of GW\,Lib as the time-tag mode of COS records detector location and arrival time of all photons. To that aim we extracted a background subtracted spectrum enclosed by a box with a cross-dispersion width of 60 pixel. We computed the background from two regions, one above and one below the spectrum, rescaled to the area of the spectral extraction region. Ly$\alpha$ at 1216\,\AA\ and \ion{O}{i} at 1302\,\AA\ airglow emission lines were masked along the spatial direction. Finally we binned the data to 10\,s resolution (Fig.\,\ref{lc}). 

While in 2002, 2010, and 2011 the system showed periodic variations on timescales of $\sim$230$-$650\,s, interpreted as non-radial white dwarf pulsations \citep[top and middle panels in Fig. \ref{lc}]{szkodyetal02-4,szkodyetal12-1}, the new observations in 2013 reveal an intriguing and puzzling feature: the light curve is dominated by a large-amplitude variability spanning the entire {\em HST} visit. 

From the observations, it is not known if the large-amplitude variability is cyclical or not; nevertheless we fitted the entire light curve with a sinusoidal function plus the 2nd and 3rd harmonics, using the Levenberg-Marquardt method for non-linear least squares minimization of the {\sc period04} package \citep{period04}. Based on the F-test, adding the 4th harmonic does not provide any significant improvement. The best fit of the fundamental period corresponds to 4.39 $\pm$ 0.09\,h and its amplitude is 175 $\pm$ 3 counts s$^{-1}$. This period is close to the $\simeq$4\,h period of the signal that has been repeatedly detected in ultraviolet and optical photometry \citep{schwieterman2010, Bullocketal11-1, Vicanetal11-1}

\begin{figure}
\centering{\includegraphics[width=1\columnwidth]{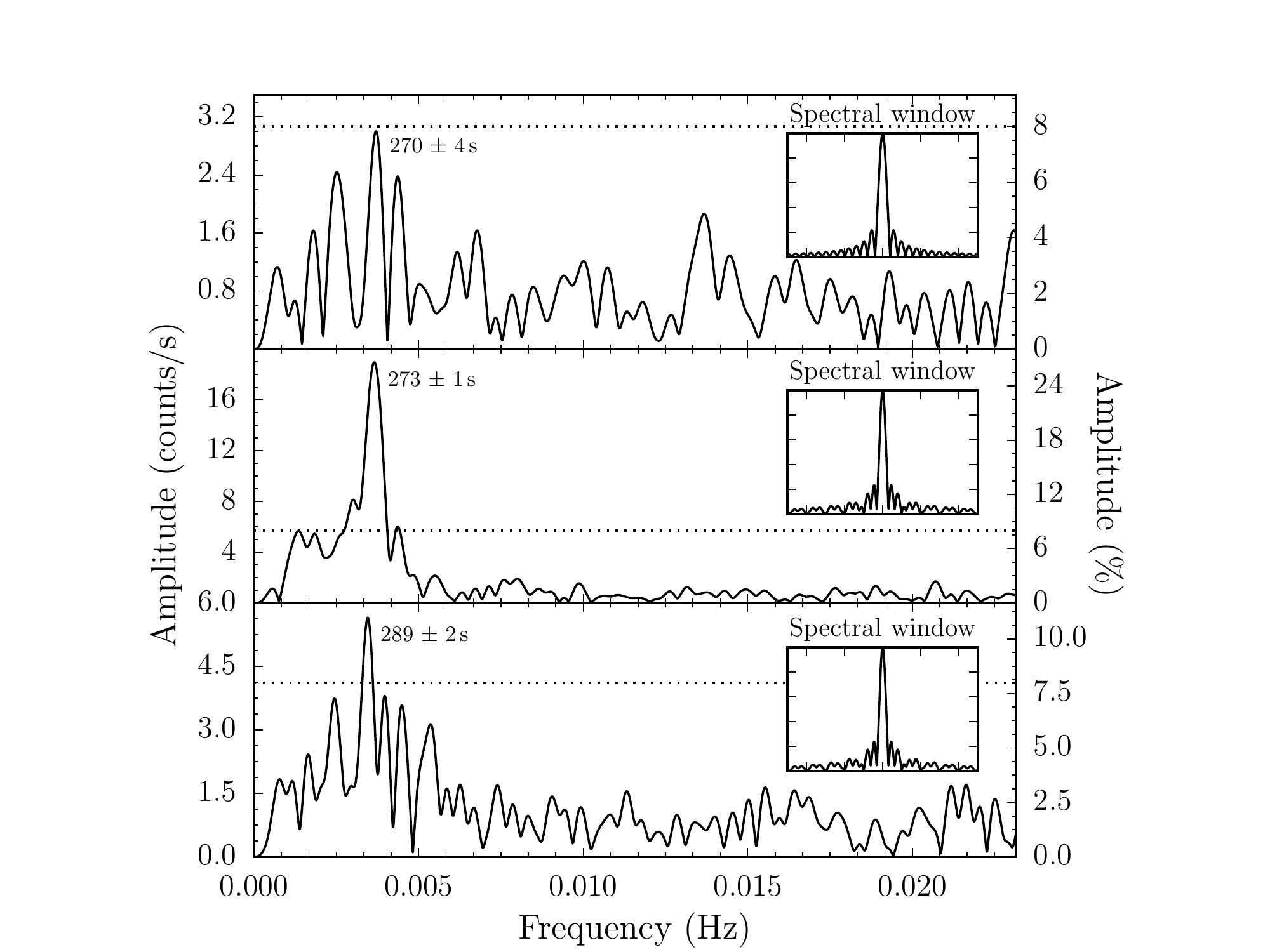}}
\caption{Power spectra for each orbit of the 2013 {\em HST} observations. The Discrete Fourier Transforms were computed after subtracting the long period variability with a polynomial fit. In each orbit the strongest signal is in the range $270-290$\,s, with a significant detection in the second and third orbit, and just falling short of the detection threshold in the first orbit. The significance level (dotted line) was defined as four times the mean of the amplitude calculated from 0 up to the Nyquist frequency.\label{dft}}
\end{figure}

In addition to the large-amplitude variability of the ultraviolet flux, the light curve shows periodic fluctuations with the highest amplitude during the second orbit (bottom panel of Fig.\,\ref{lc}). We performed a Discrete Fourier Transform using the {\sc period04} package on the light curve for each {\em HST} orbit after subtracting the long-term trend, larger amplitude variability, by fitting a second order polynomial independently to each orbit (Fig.\,\ref{dft}). In each orbit the strongest peak is found in the range 270$-$290\,s, with a statistically significant detection in the second and third orbits at 273\,$\pm$\,1\,s and 289\,$\pm$\,2\,s, respectively. We adopted the significance as four times the mean of the amplitude. Both periods are close to the 293\,s period detected in 2010 and 2011, so we suggest these periodicities correspond to white dwarf pulsations.

\section{Spectral fitting}
\label{method}

\subsection{2002, 2010, and 2011 observations}
\label{obsprev}

The excitation of gravity modes in a white dwarf is primarily a consequence of some heat flux being converted into mechanical motion in the H (He) ionisation zones, which could be either produced by an increase of the opacity compressing the overlying layers ($\kappa$-mechanism; \citealt{Dziembowski1981-1}), or more likely, the increase of opacity leads to the generation of a convecting zone (Brickhill's ''convective driving'' mechanism; \citealt{brickhill-83}, \citealt{wu+goldreich-99}). The pulsations cause geometrical distortions in the white dwarf, leading to changes in the surface gravity, which are  however, too small to be measured. The dominant effect of the pulsations is the appearance of hot and cool patterns on the white dwarf surface \citep{Robinsonetal95-2, clemensetal-2000}. The three pulsations with well-defined periods  identified during quiescence (2002 data), as well as the 293\,s period found post-outburst (2010 and 2011 data), are generally believed to be due to non-radial white dwarf pulsations. Therefore we investigated the difference in temperature produced by these pulsations. To that aim, we performed fits to two spectra, for each of the 2002, 2010, and 2011 observations, generated from the time-tag photon files. One spectrum was constructed using the photons corresponding to the sections of the light curve with the highest count rate (blue dots in Fig.\,\ref{lc}) and a second spectrum from the sections with the lowest count rate (red dots in Fig.\,\ref{lc}), hereafter referred to as \textit{peak} and \textit{trough} spectra. The thresholds of the high and low count rate are defined as a percentage above or below the mean for the full {\em HST} visit in 2002 and 2010, while for the 2011 data, the mean was calculated for each individual orbit (since the second orbit has a higher mean level, see right middle panel in Fig.\,\ref{lc}). The percentages were chosen to use only as much of the peak and trough data as to achieve an acceptable signal-to-noise ratio in the resulting spectra (7, 5, and 6 per cent for the 2002, 2010, and 2011 data, respectively).

After defining the count rate thresholds, the peak and trough STIS and COS spectra were obtained by splitting, reducing, flux-calibrating and combining the time-tag files using a series of PyRAF routines from the STSDAS task package and modules of the {\sc stsci$\_$python2.15} library.

For the spectral fits, we used a grid of white dwarf models generated with the latest version of {\sc tlusty204n} and {\sc synspec49t} \citep{hubeny+lanz95-1}. The grid covers $T_{\rm eff}$ = 9\,000$-$69\,900\,K in steps of 100\,K, $\log{g}$=8.35 dex, and a metallicity of 0.1 times the solar metallicity, which reproduces well the metal absorption lines. The models cover the wavelength range 1000$-$1800\,\AA\ with a spectral resolution of 0.094\,\AA. We performed a bilinear interpolation of the grid in wavelength and effective temperature. We fixed $\log{g}$ at 8.35, which, using the white dwarf mass-radius relation derived from cooling models for white dwarfs \citep{holberg+bergeron06-1, kowalski+saumon2006, bergeronetal2011, Tremblayetal11-1}, corresponds to previous estimates of the white dwarf mass in GW\,Lib based on the observed gravitational redshift ($M_{\rm WD}$$=$0.84$\pm$0.02\,M$_{\rm \odot}$; \citealt{vanSpaandonketal2010-2}, $M_{\rm WD}$$=$0.79$\pm$0.08\,M$_{\rm \odot}$; \citealt{szkodyetal12-1}).

The core of the broad photospheric Ly$\alpha$ absorption line shows evidence of an additional flat and featureless continuum component. Previous studies have modelled this additional continuum component by either a power-law, a blackbody spectrum or as flat in $F_{\lambda}$ \citep{Szkodyetal10-2, szkodyetal12-1, gaensickeetal05-2} and found that the exact choice of the model for the second component does not significantly affect the white dwarf parameters derived from the fit. We adopted in our analysis a flux component constant in $F_{\lambda}$, as it reduces the total number of free parameters and fit the peak and trough spectra for the effective temperature ($T_{\rm eff}$), the scaling factor ($S$), and $k$, the flux of the constant $F_{\lambda}$ component.

\begin{table*}
\begin{center}
\caption[Fitting Parameters]{Best-fit parameters obtained using the affine-invariant ensemble sampler for MCMC, with a fixed $\log{g}$ =8.35. Uncertainties are the 1-$\sigma$ confidence interval resulting from flat priors on $T_{\rm eff}$, $S$ and $k$. White dwarf radii ($R_{\rm WD}$) are computed from the flux scaling factor adopting a distance of 104$^{+30}_{-20}$ pc \citep{Thorstensen03-3}.  \label{tab:teff}}

\begin{tabular}{c c c c }
\hline
\hline

Parameter & Peak spectrum  & Average spectrum  & Trough spectrum\\
\hline

$T_{\rm eff, 2002}$ (K)    &    14\,918$^{+18}_{-20}$     &	14\,695$^{+13}_{-11}$     &	14\,440$^{+22}_{-22}$     \\
$T_{\rm eff, 2010}$ (K)    & 	18\,343$^{+62}_{-69}$     &	17\,980$^{+14}_{-14}$     &	17\,872$^{+75}_{-72}$     \\	
$T_{\rm eff, 2011}$ (K)    & 	16\,160$^{+62}_{-61}$	  &	15\,915$^{+ 9}_{- 9}$	  & 15\,748$^{+62}_{-61}$	  \\
\hline
$R_{\rm WD, 2002}$ (R$_{\rm \odot}$) & 	0.015$^{+0.009}_{-0.005}$ &	0.014$^{+0.009}_{-0.005}$ &	0.015$^{+0.009}_{-0.005}$ \\
$R_{\rm WD, 2010}$ (R$_{\rm \odot}$) & 	0.017$^{+0.010}_{-0.006}$ &	0.016$^{+0.010}_{-0.006}$ &	0.016$^{+0.010}_{-0.006}$ \\
$R_{\rm WD, 2011}$ (R$_{\rm \odot}$) &	0.018$^{+0.010}_{-0.006}$ &	0.018$^{+0.010}_{-0.006}$ &	0.018$^{+0.011}_{-0.006}$ \\
\hline
$k_{\rm 2002}$ ($\times 10^{-16}$ erg\,s$^{-1}$\,cm$^{-2}$\,\AA$^{-1}$) &4.5$^{+0.2}_{-0.2}$ & 3.92$^{+0.07}_{-0.07}$ & 2.7$^{+0.1}_{-0.1}$ \\
$k_{\rm 2010}$ ($\times 10^{-15}$ erg\,s$^{-1}$\,cm$^{-2}$\,\AA$^{-1}$) &2.8$^{+0.2}_{-0.2}$ & 3.22$^{+0.05}_{-0.05}$ & 2.4$^{+0.2}_{-0.2}$\\
$k_{\rm 2011}$ ($\times 10^{-15}$ erg\,s$^{-1}$\,cm$^{-2}$\,\AA$^{-1}$) &1.2$^{+0.1}_{-0.1}$ & 1.30$^{+0.02}_{-0.02}$ & 1.2$^{+0.1}_{-0.1}$\\

\hline
\end{tabular}
\end{center}
\end{table*}

\begin{figure}
\centering{\includegraphics[width=1\columnwidth]{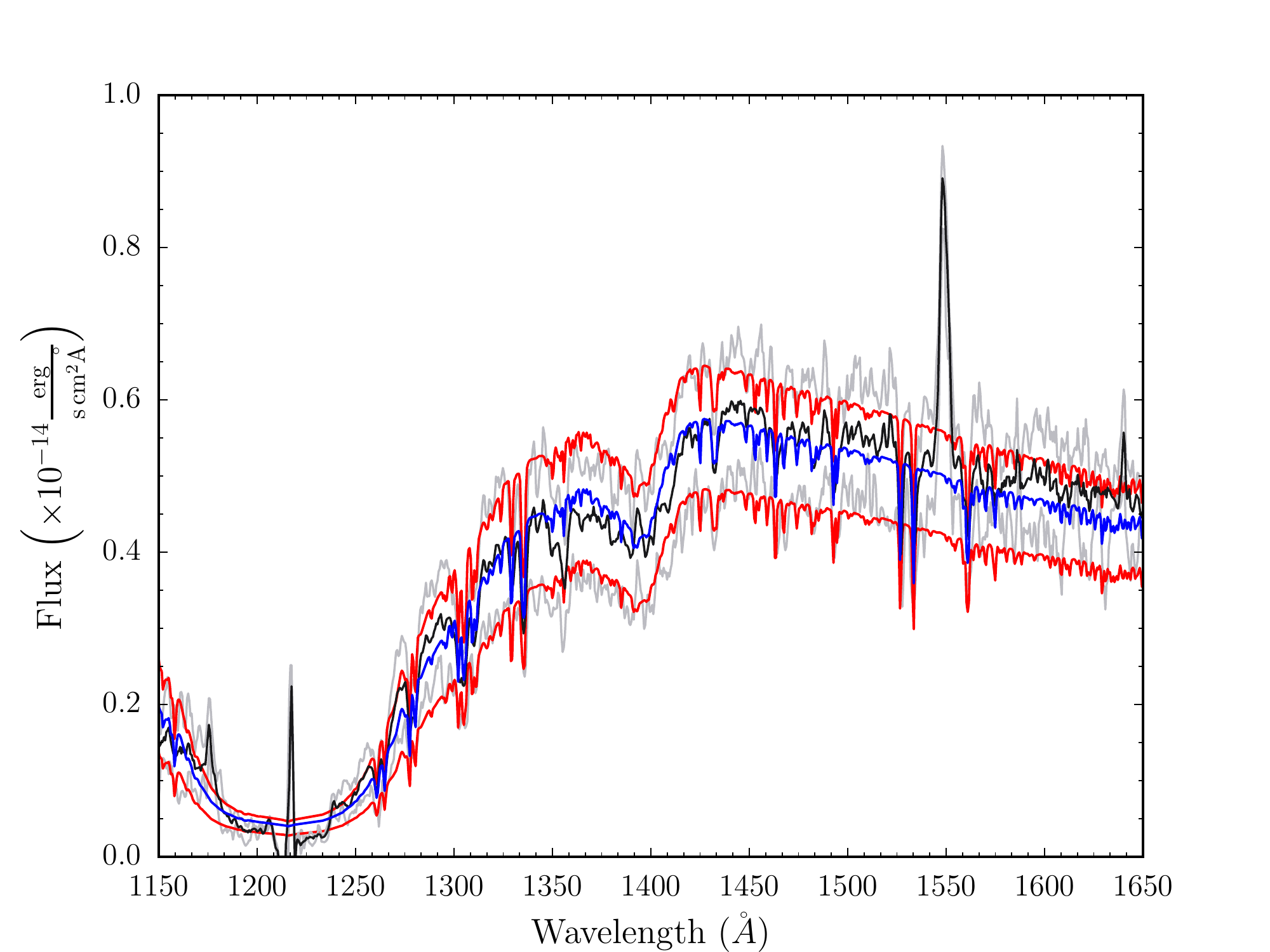}}
\caption{The black solid line represents the average spectrum for the 4 orbits of the STIS data taken in 2002. The blue line corresponds to the best MCMC-fit with $T_{\rm eff}$=14\,695$^{+13}_{-11}$\,K. The upper and lower gray lines show the peak and trough spectra, respectively (see Section \ref{obsprev} and Fig. \ref{lc}). The red lines are the best-fit models, $T_{\rm eff}$=14\,918$^{+18}_{-20}$\,K (peak) and $T_{\rm eff}$=14\,440$^{+22}_{-22}$\,K (trough). \label{stis}}
\end{figure}

We used the Markov Chain Monte Carlo (MCMC) EMCEE open-source code implemented in Python \citep{emcee}. We constrained $T_{\rm eff}$ with a flat prior function over the range 10\,000$-$20\,000\,K, based on previous measurements \citep{szkodyetal02-4, szkodyetal12-1}, as well as $k$ and $S$ to be positive. Finally we defined the log-likelihood function to be $-\chi^{2}$/2. To estimate an initial guess for the parameters we used the Levenberg-Marquardt minimization method. We masked the Ly$\alpha$ airglow line and the \ion{C}{iv} emision line during the fit, using the ranges of 1208$-$1223\,\AA\ and 1543$-$1555\,\AA, respectively.

In general, the MCMC samples were well burnt-in before 100 iterations, and the one-dimensional projection of the posterior probability distributions of the parameters follow a Gaussian distribution. Therefore we chose the median of this distribution to be the best-fit value for the given parameter, and defined the internal uncertainty as calculated based on the 15.87 and 84.1 percentiles of the posterior probability distribution.

The best fit effective temperatures for the peak and trough spectra of 2002, 2010, and 2011 are listed in Table \ref{tab:teff}. The results show that the pulsations led to a difference of nearly 500\,K over the visible surface of the white dwarf. This temperature difference is clearly noticeable between the peak and trough spectra, as shown in Fig. \ref{stis} for the 2002 STIS data.

The white dwarf radius ($R_{\rm WD}$) can be derived using the scaling factor ($S$) and the distance ($d$) of 104$^{+30}_{-20}$ pc \citep{Thorstensen03-3},

\begin{equation}
S= \pi \left(\frac{R_{\rm WD}}{d}\right)^{2}.
\label{equation}
\end{equation}

The resulting radii are listed in Table \ref{tab:teff}, where the large uncertainties are primarily systematic in nature, resulting from the error on the distance measurement. Within the uncertainties, the derived radii agree with the expected radius of a white dwarf of mass $\simeq$0.84\,M$_{\rm \odot}$ located at the distance of GW\,Lib\footnote{The radius was derived from DA cooling models \citep[]{holberg+bergeron06-1, kowalski+saumon2006, bergeronetal2011, Tremblayetal11-1} http://www.astro.umontreal.ca/$\sim$bergeron/CoolingModels}

The measured temperature variations of a few hundred Kelvin are consistent with non-radial pulsations producing an inhomogeneous photospheric temperature distribution across the visible white dwarf hemisphere, which is reflected in flux and colour variations \citep{Fontaineetal1982}.

\subsection{2013 observation}
\label{obs2013}

While we expect that white dwarf pulsations cause the short-period oscillations seen in the light curve, particularly in the second orbit, the nature of the $\simeq$4.4\,h flux variation (bottom panel Fig.\,\ref{lc}) is unclear, as is its physical location within the CV. Here, we explore a white dwarf photospheric origin. For that purpose, we process the 2013 COS data into a set of 17 spectra, as labelled underneath the light curve in the bottom panel of Fig.\,\ref{lc}. The average exposure time for each spectrum is $\simeq$435\,s.

\subsubsection{Two-component model}
\label{2-comp}
We then fitted this set of time-resolved spectra following the same procedure as described in Sect. \ref{obsprev}. In each of the spectra the dominant ultraviolet flux component can be described by a white dwarf model. In addition, a small second continuum component, which we modelled with a constant F$_{\lambda}$ $k$, is required, as easily identified near the core of the broad photospheric Ly$\alpha$ absorption. The sequence of spectra clearly shows a smooth variation in both total flux and overall shape. The Ly$\alpha$ line is broadest in spectrum \#1 (grey line in Fig.\,\ref{spectra_2013}), becomes narrower throughout the sequence up to spectrum \#12 (black line in Fig.\,\ref{spectra_2013}), and then broadens again. We found that the observed change in the width of the Ly$\alpha$ absorption can be reproduced by an increase and subsequent decrease in the effective temperature of the white dwarf. The results are shown in blue dots in Fig. \ref{results}. The white dwarf temperature ($T_{\rm eff}$) changes from 15\,975$^{+39}_{-37}$\,K (spectrum \#5) up to 18\,966$^{+46}_{-47}$\,K (spectrum \#12) which correlates with the constant flux ($k$) and anticorrelates with the \textit{apparent} white dwarf radius (R$_{\rm APP}$). We adopt the designation of apparent radius as a change in the actual white dwarf radius is unphysical. We interpret these results as heating of a localised region on the white dwarf. This region dominates the ultraviolet flux, resulting in a decrease of the apparent radius.

\subsubsection{Three-component model}
\label{3-comp}
To confirm our assumption of a hotter region on the white dwarf surface, we repeated the fits, however in addition to the global white dwarf model and the second continuum component, $k$, we included a second, hotter, white dwarf model to model this region. The area covered by this hot region was a free parameter, and the temperature of the global white dwarf was fixed to 15\,975\,K, which corresponds approximately to the temperature of the coolest spectrum (labelled as \#5 in Fig. \ref{lc}). To avoid a temperature inversion between the hot region and the cool underneath white dwarf, we constrain the temperature of the hot spot to be higher than $T_{\rm eff}$=15\,975\,K

\begin{figure}
\centering{\includegraphics[width=1\columnwidth]{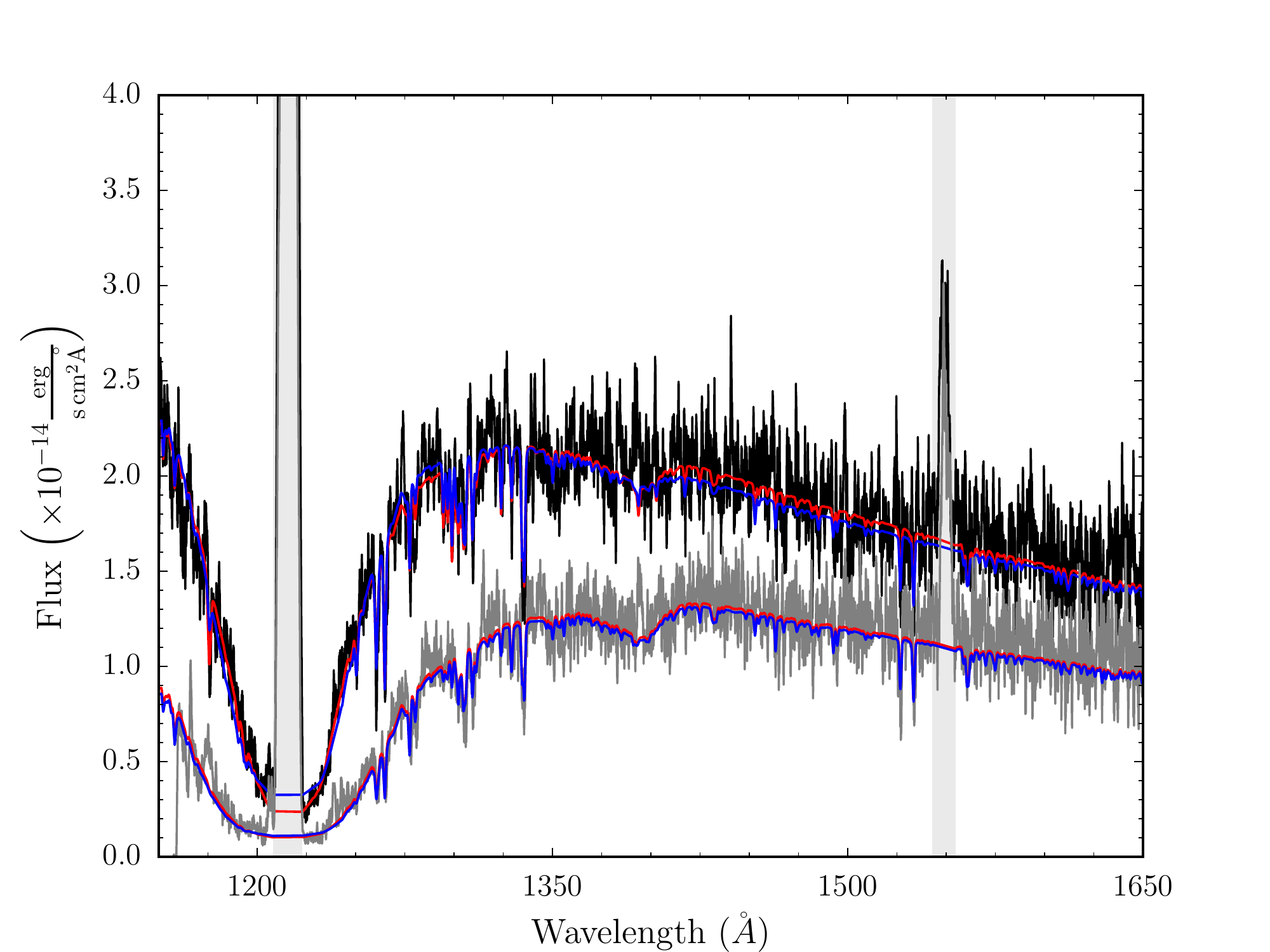}}
\caption{The grey and black lines display the spectra \#1 and \#12 defined in Fig.\,\ref{lc}. We overplotted the best-MCMC fits using two-component model (blue) and three-component model (red). The spectrum \#1 is well-fitted by a single white dwarf with $T_{\rm eff}$=16\,102$^{+37}_{-37}$\,K, while spectrum \#12 is better fitted by the three component model, particularly in the central region of the Ly$\alpha$ absorption, using a global white dwarf at $T_{\rm eff}$=15\,975\,K plus a hotter (22\,060$^{+313}_{-303}$\,K) white dwarf model covering $\simeq$31 per cent of the visible white dwarf surface. The airglow line of Ly$\alpha$ and \ion{C}{iv} (shaded in grey) were masked during the fit. The spectra were smoothed with a 5-point boxcar for display purposes. \label{spectra_2013}}
\end{figure}

\begin{figure}
\centering{\includegraphics[width=1\columnwidth]{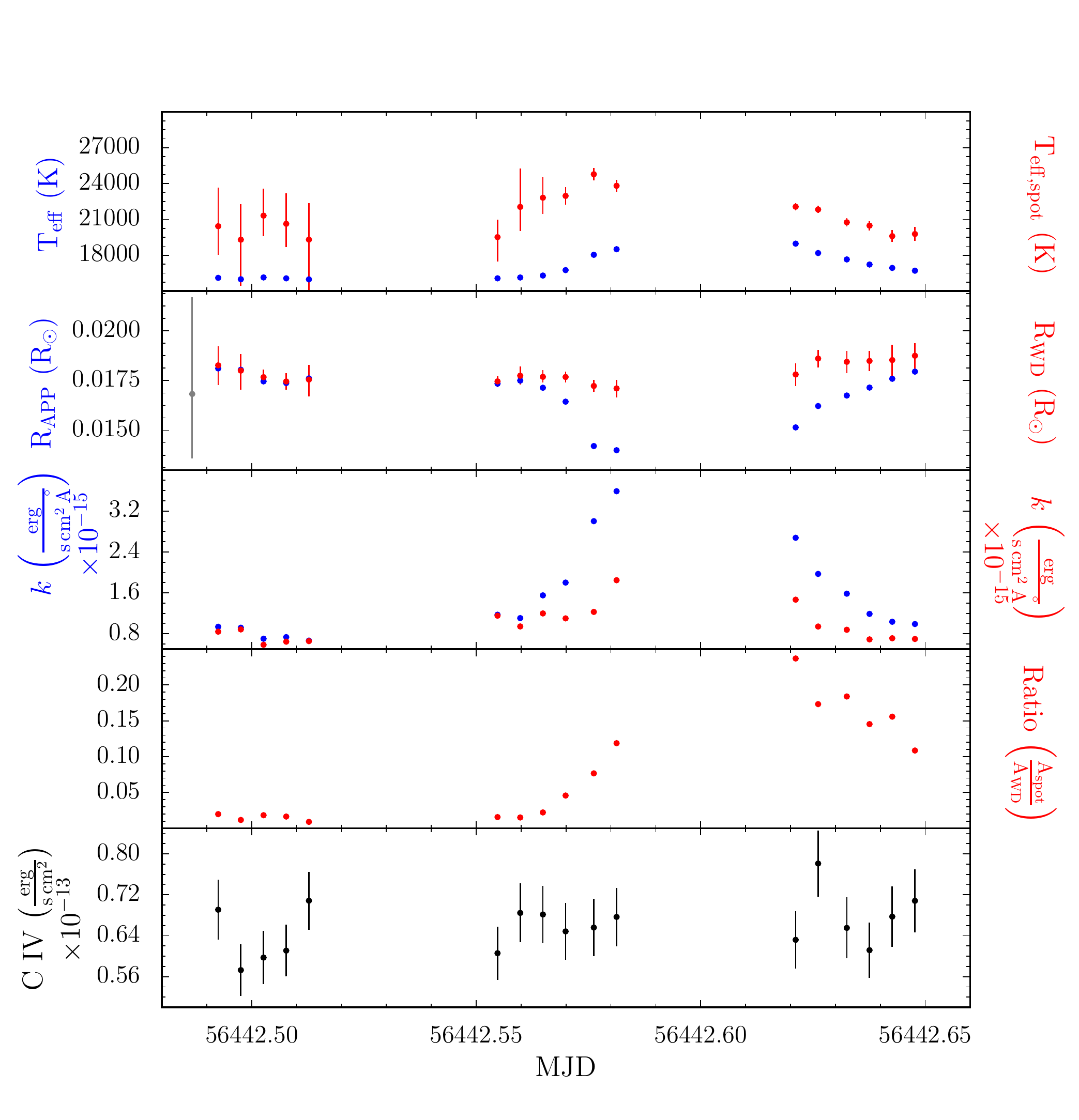}}
\caption{Best-fit parameters for the set of 17 spectra processed from the May~2013 observations. We fitted the set using a two-component model (blue dots): a white dwarf plus a flat F$_{\lambda}$ ($k$) component, and with a three-component model (red dots): an underlying white dwarf at $T_{\rm eff}$=15\,975\,K, a flat F$_{\lambda}$ component ($k$), and a hotter white dwarf accounting for a localised heated fraction of the white dwarf. The intrinsic uncertainties have been estimated based on the 15.87 and 84.1 percentiles of the MCMC samples in the marginalized distributions, but some of them are too small to be seen on this scale. The distance measurement used to determine the apparent white dwarf radius introduces a large systematic uncertainty (gray error bar). This systematic uncertainty affects, however, the entire sequence of apparent radii in the same way, i.e. the shape of the apparent radius variation is a robust result. The fourth panel shows the ratio between the fraction covered by the hotter white dwarf and the area of the global white dwarf. The fifth panel shows the flux of \ion{C}{iv} emission line at 1550\,\AA. See Section \ref{discussion} for details. \label{results}}
\end{figure}

The best-fit parameters are illustrated in red dots in Fig. \ref{results}. The temperature of the hot region ($T_{\rm eff, spot}$) on the white dwarf increases from the global white dwarf temperature of 15\,975\,K to 24\,782$^{+506}_{-521}$\,K. The white dwarf radius is calculated using equation \ref{equation}, where $S$ is the sum of the scaling factors of the $T_{\rm eff}$=15\,975\,K underlying white dwarf and the hot region. We found that the white dwarf radius remains between 0.016$-$0.018R$_{\rm \odot}$, which, we note, it is consistent with the radius estimates from the 2002, 2010, and 2011 observations (Table \ref{tab:teff}). In contrast to the two-component fit in Section \ref{2-comp}, the additional flat F$_{\lambda}$ $k$ component no longer shows a strong correlation with the temperature of the hot region, as the region of the core of Ly$\alpha$ can be better reproduced with the inclusion of a hotter white dwarf (Fig. \ref{spectra_2013}). In the fourth panel we show the ratio between the area of the hot region and the total visible white dwarf area. It shows that during the first seven spectra the area of the hot region is practically non-existent, explaining the large errors in the temperatures of the hot region. Once the ultraviolet flux rises and the area of this region grows, its temperature rises rapidly. Additionally, we show the flux of the \ion{C}{iv} emission line in the bottom panel, which we discuss below. 

\section{DISCUSSION}
\label{discussion}

\subsection{Possible scenarios explaining the change in flux of the white dwarf}
In Section \ref{obs2013} we demonstrated that the evolution of both the total ultraviolet flux, as well as the morphology of the Ly$\alpha$ profile obtained in 2013, is well-described by an increase of a few 1000\,K, and subsequent decrease, in the temperature of a fraction of the white dwarf surface. The area of this hot region covers up to $\simeq$30 per cent of the visible white dwarf area. We found this variation lasts $\simeq$4.4\,h which is close to the  recurrent $\simeq$4\,h period signal reported previously \citep{schwieterman2010, Bullocketal11-1, Vicanetal11-1}. The three groups found that the signal wanders in phase and amplitude on time scales of days and occasionally disappears. In summary, a $\simeq$4\,h variability that is detected both in the optical and ultraviolet appears to be a recurrent feature in GW\,Lib and our analysis of the new {\em HST} observations link it to an apparent heating and cooling of the white dwarf. Here we discuss possible scenarios that could cause such a change in the white dwarf temperature.

\subsubsection{An accretion spot on a magnetic white dwarf.}

Accretion causes heating of the white dwarf, and non-uniform accretion of matter will result in an inhomogeneous temperature distribution over the white dwarf surface. Accretion-heated spots that exceed the temperature of the underlying white dwarf by several 1000\,K are frequently observed in polars, i.e. strongly magnetic CVs, and cause a modulation of the ultraviolet flux and the width of the Ly$\alpha$ profile on the spin period of the white dwarf \citep{gaensickeetal95-1, gaensickeetal98-2, gaensickeetal06-2, schwopeetal02-1, araujo-betancoretal05-2, szkodyetal10-1}. Assuming the presence of a weak magnetic field (B) we can calculate the minimum magnitude required by the white dwarf to decouple the electron spin and detect splitting of sharp metal lines with the spectral resolution of COS. In the case of \ion{Si}{ii} the absorption at 1260\,\AA\ and 1265\,\AA\ corresponds to electron transitions  of $^{2}P_{1/2} \rightarrow ^{2}D_{3/2}$ and $^{2}P_{3/2} \rightarrow ^{2}D_{5/2}$, respectively. Under a weak magnetic field regime, the $^{2}D_{5/2}$ and $^{2}D_{3/2}$ levels of the \ion{Si}{II} would split into six and four states, respectively. We found that necessarily $B > 1.4$\,MG for $^{2}D_{3/2}$ (Land\'e factor equals to 0.8) and $B>0.9$\,MG for $^{2}D_{5/2}$ (Land\'e factor equals to 1.2) to see such splitting. We conclude that, because we do not detect Zeeman splitting in the \ion{Si}{II} at 1260\,\AA, the magnetic field of the white dwarf in GW\,Lib is limited to $B\le0.9$\,MG. However, in polars, the rotation of the white dwarf is locked to the orbital period, in the range $\simeq$1.5$-$8\,hr, whereas the spin period of GW\,Lib is $\simeq$100$-$200\,s \citep{vanSpaandonketal2010-2, szkodyetal12-1}. Hence, accretion onto a magnetic white dwarf would result in coherent variability on time scales of $\simeq$100$-$200\,s, and we therefore rule out this scenario to explain the $\simeq$4\,h variability.

\subsubsection{A brief increase in the accretion rate.}

Another possible scenario explaining the observed $\simeq$4\,h variability are quasi-periodic brief accretion events that significantly heat a fraction of the white dwarf. However, such intermittent heating needs to be extremely symmetric with respect to the spin axis of the white dwarf, as we do not observe any variability of the ultraviolet flux at the white dwarf spin period. Accretion-heating of an equatorial belt of a non-magnetic white dwarf \citep{kippenhahn+thomas78-1} would match this constraint well.

\begin{figure}
\centering{\includegraphics[width=1\columnwidth]{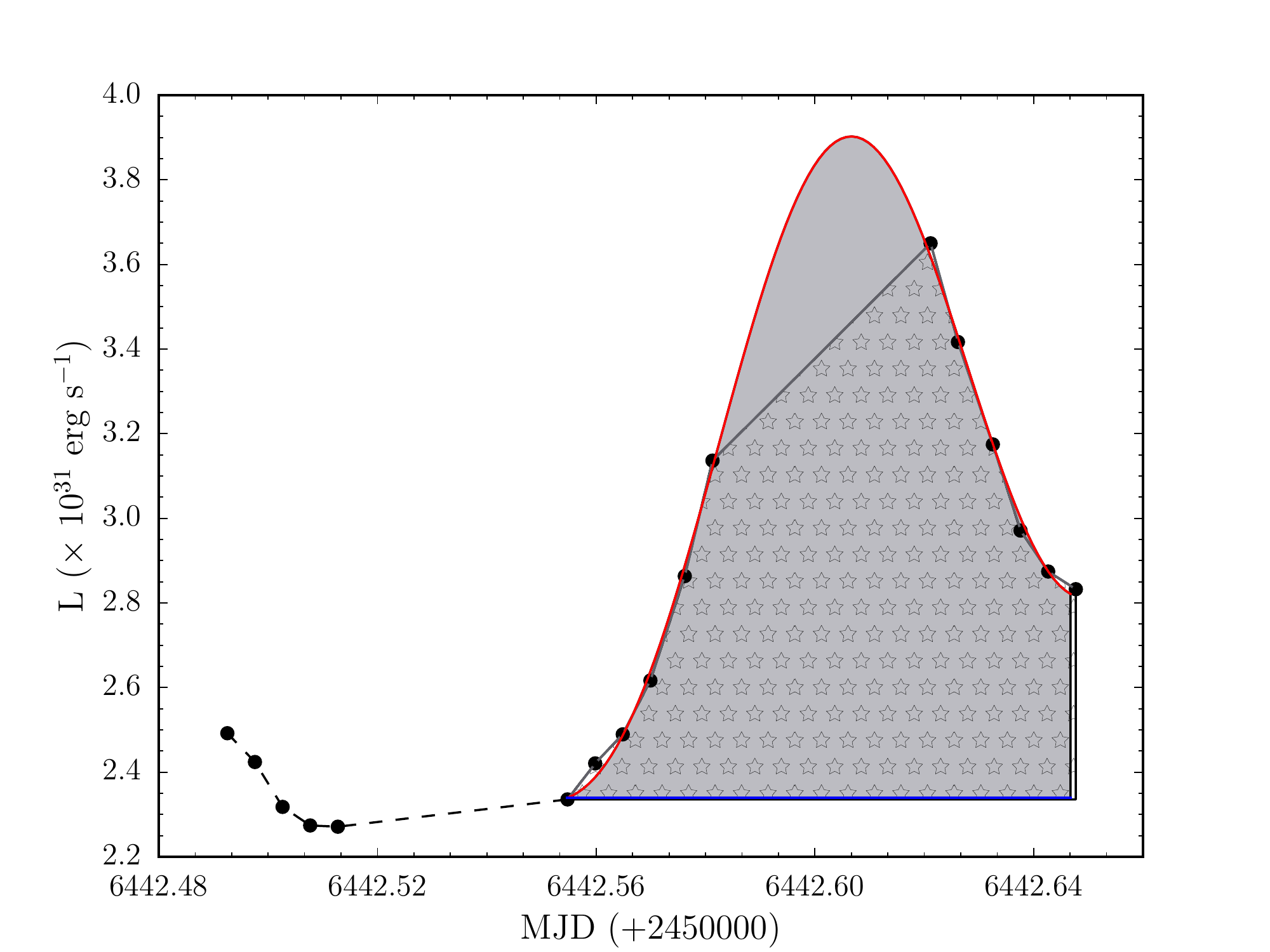}}
\caption{Luminosity as a function of time calculated from integrating the flux of white dwarf models with the effective temperatures given in Fig \ref{results} (blue dots). The energy excess is calculated using a radius of 0.018\,R$_{\rm \odot}$ and a mass of 0.84 M$_{\rm \odot}$ for the white dwarf, and subtracting the luminosity of the underlying white dwarf model with $T_{\rm eff}$=16\,063\,K (area above the blue line). The starred area represents an excess of energy of 6.2$\times 10^{34}$\,ergs while the grey area represents 7.1$\times 10^{34}$\,ergs of energy excess. \label{lum}}
\end{figure}

We can estimate the excess of energy released during a hypothetical accretion episode from the excess luminosity. We derived the luminosity by integrating white dwarf models that follow the temporal evolution of the effective temperature shown as blue dots in the top panel of Fig. \ref{results}, scaled to the observed flux and subtracting the flat component (Fig. \ref{lum}). The energy excess corresponds to the area above of the luminosity of the white dwarf with $T_\mathrm{eff}=16\,063$\,K, corresponding to spectrum \#6 (blue line in Fig. \ref{lum}) and, adopting a white dwarf mass of 0.84\,M$_{\rm \odot}$ \citep{vanSpaandonketal2010-2} and the apparent radius of 0.018\,R$_{\rm \odot}$ (See section \ref{obs2013}), we can derive a total mass accreted during the accretion episode. The discontinuity of the data, specifically near the peak of the $\simeq$4.4\,h variability, does not allow us to accurately define the shape of the area to estimate the energy excess. The simplest interpolation is connecting the points with a straight line, which gives an energy excess of 6.2$\times 10^{34}$\,ergs (starred area), and a total mass of $\simeq$ 3.5$\times10^{-16}$\,M$_{\rm \odot}$, to be accreted during a hypothetical accretion episode that lasts $\simeq$2.2\,hr. However this estimate is strictly a lower limit. As an alternative, we fitted the luminosity adopting the same model we used to fit the light curve (i.e. sinusoidal function plus 2nd and 3th harmonics, red solid line). This approach gives an energy excess of 7.1$\times10^{34}$\,ergs (grey area) and, consequently, an accreted mass of 4.0$\times10^{-16}$\,M$_{\rm \odot}$ in $\simeq$2.2\,h, i. e. the two approaches do not differ significantly, and suggest that a brief accretion episode with 1.3$-$1.5$\times10^{-12}$M$_{\rm \odot}$/yr could cause the observed heating. This rate is very low compared to the expected accretion rate for DNe in quiescent states \citep{townsley+gaensicke09-1, goliasch+nelson15}.

In principle, small changes in the accretion rate onto the white dwarf could be caused by efficient irradiation of the innermost region of the disc by the slowly cooling white dwarf, keeping these regions in an ionised state while the outer disc is cooler and neutral. Instabilities in the transition region could cause quasi-periodic fluctuations of the accretion rate, resulting in heating and subsequent cooling of the white dwarf. The time scale of such cycles would depend very much on the temperature of the white dwarf.

Another possibility is that magnetic field star-disc interactions cause quasi periodic changes of the accretion rate in the inner disc. According to \citet{Uzdensky2002-1}, the connection between field lines of a disc and star can be periodically broken due to twisting and reconnection. One may speculate that such cycles can generate quasi-periodically occurring episodes of enhanced accretion. 

If the $\simeq$4\,h variability seen in GW\,Lib is indeed related to quasi-periodic brief accretion events, one would expect similar activity in additional CVs. \citet{woudt+warner02-2} and \citet{Vicanetal11-1}  note that only a handful of other systems show quasi-periodic variability at periods significantly longer than their orbital periods, in particular FS\,Aur \citep{tovmassianetal03-1}. Although quasi-periodic accretion events are in principle a possible explanation for our 2013 COS observations of  GW\,Lib, we consider this scenario unlikely because of two reasons. Firstly, we would expect an increase in the accretion rate onto the white dwarf to be accompanied with photometric stochastic flickering (which is a characteristic of FS\,Aur, \citealt{Neustroevetal2013-1}), which is not observed. Secondly, enhanced accretion activity should be linked to a variation of the \ion{C}{iv} emission line flux, which is also not observed (Fig. \ref{results}).

\subsubsection{Retrograde wave in a rapid rotating white dwarf}

\citet{schwieterman2010} speculated about unusually long pulsations modes in the white dwarf as a possible origin of the $\simeq$4\,h variability. In single, slowly-rotating white dwarfs, pulsation periods can be as long as $\simeq$20\,min (e.g. \citealt{Mukadametal06-1}); modes with periods of several hours are physically unlikely. 

\begin{figure}
\centering{\includegraphics[width=1\columnwidth]{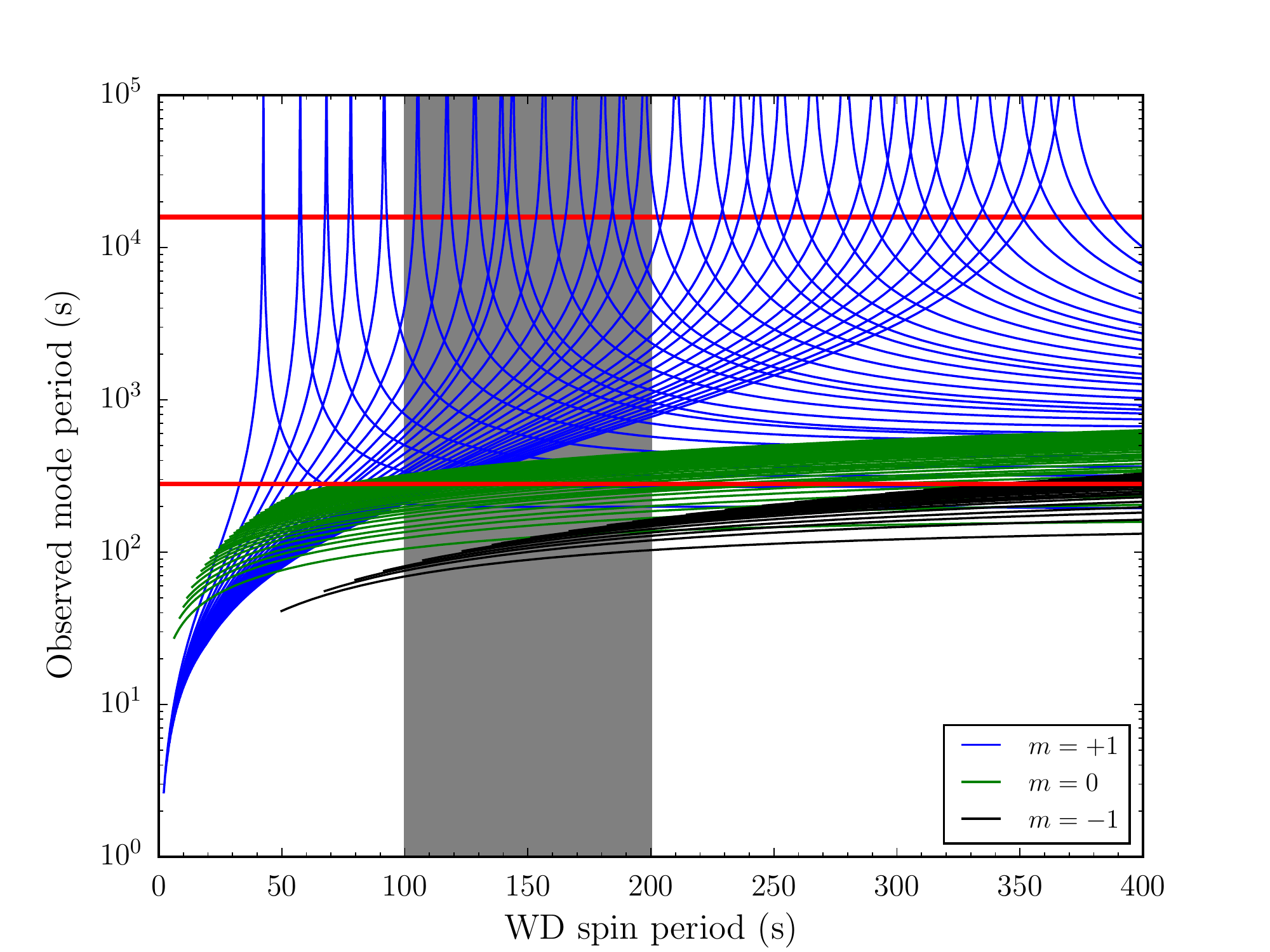}}
\caption{Observed mode periods produced by the splitting of the \textit{g}-modes due to fast rotation of the white dwarf \citep{townsleyetal-16}, where $m=1$ (retrograde, blue), $m=-1$ (prograde, black), and $m=0$ (green). This model was computed using a total mass, $M=0.9$M$_{\rm \odot}$, a H-rich accreted layer mass of half of the nova ignition mass (calculated as described in \citealt{townsley+bildsten-2004}), and $T_{\rm eff}$ $=15\,500$\,K \citep{szkodyetal02-4}. The red bands show the $\simeq$4.4\,h-period variability (top) and the $\sim$280\,s pulsation signal (bottom) identified in the 2013 observations. The light shade area represents the white dwarf spin period \citep{vanSpaandonketal2010-2, szkodyetal12-1}.  \label{pulsations}}
\end{figure}

However, the white dwarf in GW\,Lib rotates extremely rapidly with a spin period of $\simeq$100$-$200\,s \citep{vanSpaandonketal2010-2, szkodyetal12-1} compared to the slow spin periods of hours to days in single white dwarfs \citep{kawaler2004}. In the case of rapid rotation, $\Omega_{\rm spin} > \Omega_{\rm mode}$, the "splitting" of \textit{g}-modes by rotation is no longer a small perturbation, and the mode period can change by order unity or more.  Also, modes on rotating stars are sensitive to the rotation axis and the direction of rotation.  As a result, for $l=1$ modes, the modes that corresponds to the $m=-1$ spherical harmonics propagate in a prograde direction (around the star in the same direction as the spin) while the $m=+1$ mode corresponds to retrograde modes.  For the retrograde modes, the frequency measured by a fixed observer is $\Omega_{\rm obs} = |\Omega_{\rm mode} - \Omega_{\rm spin}|$.  The \textit{g}-mode spectrum will be a sequence of modes starting at around a few hundred seconds and extending to longer periods as the radial order of the mode increases \citep{Unnoetal-1989, bildstenetal1996}.  As a result, for a star like GW\,Lib, which is rotating with a period of around 100$-$200 seconds, there can be a low-radial-order \textit{g}-mode, $n\sim 5$, for which $\Omega_{\rm mode} \approx \Omega_{\rm spin}$ so that $\Omega_{\rm obs}$ is much closer to zero, thus giving a significantly longer period.  This is a mode that, by propagating opposite to rotation on the star, ends up being almost fixed in from the observers' point of view.  The excitation of this mode could give rise to the observed $\simeq$4.4\,h period (Fig. \ref{pulsations}). 

The model white dwarf shown in Fig. \ref{pulsations} has parameters (mass, accreted layer, accretion rate, $T_{\rm eff}$) similar to those expected for GW\,Lib, but is not a fit \citep{townsleyetal-16}.  The model demonstrates that an accreting white dwarf rotating at the spin rate observed for GW\,Lib will have low to moderate order modes that are consistent with both the observed $\sim$4.4\,h mode as well the $\sim$280\,s mode.  The low order modes are those with shortest mode period in the slowly rotating GW\,Lib, periods at the right edge of the figure.  If the white dwarf spin is near 100\,s, the best candidate mode in this model for the $\sim$4.4\,h period is the retrograde ($m=+1$) $n=6$ radial order mode.  In this case the $\sim$280\,s period may correspond to either a lower radial order (shorter period, e.g.  $n=2$) mode that is short enough period to still be retrograde in the observer's frame, or one or several high order retrograde modes that have periods sufficiently longer than the spin period that the spin of the white dwarf has "dragged" them be prograde from the point of view of the observer.  White dwarf models with slightly different parameters will have normal modes that are shifted up or down by up to of order 100 seconds, that will change which radial order mode matches the $\sim$4.4\,h period.

Unlike the spherically symmetric or low-spin case, rotationally modified \textit{g}-modes do not extend as evenly over the entire surface of the star.  Their amplitudes are larger near the equatorial regions \citep{bildstenetal1996}.  The equatorial band over which the mode's amplitude is large depends on the ratio between $\Omega_{\rm mode}$ and $\Omega_{\rm spin}$, with lower frequency (higher radial order) modes being constrained more tightly near the equator, and higher frequency (lower radial order) having more extension to higher latitude. The moderate radial order necessary to have $\Omega_{\rm spin}\approx\Omega_{\rm mode}$ could therefore also lead to the decrease in the effective area as the mode contributes more to the flux. Some of this contribution may be blocked by the edge-on accretion disk, but if the mode extends to high enough latitude to be visible it would lead to a reduction in area like that observed.

We note that rare large-amplitude brightening episodes have also been observed in two pulsating white dwarfs. PG\,1149+057 \citep{Hermesetal-2015} and KIC\,4552982 \citep{Belletal-2015} exhibit recurrent outbursts lasting 4$-$40\,h longer than the pulsation periods identified in these systems. PG\,1149+057 and KIC\,4552982 brighten overall by up to 45\% and 17\%, respectively, not quite as large as the observed amplitude of the 4.4\,h variability in GW\,Lib. The recurrence time of these outbursts is $\sim$days and, when such an event occurs, the pulsation spectrum increases in amplitude supporting the fact that the outbursts originate on the white dwarf. \citet{Hermesetal-2015} suggest that these events are most likely related to white dwarf pulsations.

There are two possible reasons that the mode currently responsible for the $\simeq$4.4\,h variability is not continuously detected. One is simply that it was not excited. The second is that if its period shifts by a small amount it may no longer be similar enough to the spin period to lead to long time scale variability in the observer frame. While large shifts in periods are not expected due to the stability of the overall white dwarf structure, a small shift that may arise from the heating and cooling of the outer layer due to the accretion event may be sufficient to change the match between the mode and spin period.

\subsection{Nature of the second component}

Some extra flux contribution is clearly present in the broad core of the photospheric  Ly$\alpha$ line. As outlined above, similar additional continuum flux components have been detected in many other DN systems \citep{godonetal2004-2, longetal2009-1, sionetal03-1, gaensickeetal05-2}. Under the assumption of a featureless flat F$_{\lambda}$ model, its contribution to the total flux drops from $\simeq$8\%$-$26\% to $\simeq$7\%$-$13\%, when we included a second white dwarf model representing the heated region. We find a correlation between the flux of this flat F$_{\lambda}$ component and the effective temperature of the white dwarf in the 2013 data (Fig. \ref{results}). Such correlation is not quite obvious in the {\em HST} spectroscopy taken in 2002, 2010, 2011 (Table \ref{tab:teff}), where the difference in temperature between the trough and peak spectra is only a few hundred Kelvin.

To further explore the nature of this component, we repeat the fits using the three-component model but we included the second component as optically thick thermal blackbody emission instead of a flat in F$_{\lambda}$ model, and found temperatures in the range of 18\,000$-$28\,000\,K, which mimicks in an approximately a flat F$_{\lambda}$ continuum over the wavelength range of the {\em HST}/COS data.  However, the area of this thermally emitting region is quite small, reaching at most to $\sim$4\% that of the white dwarf area. Nonetheless, a blackbody model for the second component does not improve the overall fits compared to the flat F$_{\lambda}$ model and the resulting white dwarf temperatures shown in Fig. \ref{results} remain practically unchanged.

For completeness, we repeated the fits using only single white dwarfs models. Though the white dwarf temperatures are higher by $\simeq$300\,K, the overall shape of the temperature variation follows the same trend as shown in Fig. \ref{results}. However these fits completely fail to reproduce the observed core of the Ly$\alpha$ absorption line.

While we cannot unambiguously identify the physical nature of the second component, incorporating either a flat F$_{\lambda}$ continuum or a blackbody significantly improves the fits compared to using only single white dwarf models.

\section{Conclusion}
\label{conclusions}

We have presented the analysis of new and archival {\em HST} observations of the DN GW\,Lib taken in 2013, whose UV flux in quiescence is known to be dominated by the white dwarf, and which is known to exhibit non-radial mode pulsations. 

We have analysed ultraviolet {\em HST} observations obtained to pre-outburst in 2002, and post-outburst in 2010, and 2011, demonstrating that in fact, the non-radial pulsating signals identified in these observations lead to variations of a few hundred Kelvin over the visible white dwarf surface.

We identified the presence of short period oscillations in the new 2013 observations, with a significant detection during the the second and third orbits at periods of $\simeq$273\,s and $\simeq$289\,s, respectively. The amplitude is strongest in the second orbit. We suggest these oscillations are produced by pulsations of the white dwarf.

We also identified a large amplitude variability spanning the entire COS observations, with a long period of $\simeq$4.4\,h, and we suggest to be the same $\simeq$4\,h signal previously detected in the ultraviolet and optical \citep{Bullocketal11-1, schwieterman2010, Vicanetal11-1}. We demonstrated that this $\simeq$4.4\,h variability can be explained by a simultaneous increase of the white dwarf temperature and a decrease of its apparent radius. Subsequently, the white dwarf cools while the apparent radius is relaxing back to its original size. We postulate that this large temperature change occurs only over a fraction of the white dwarf surface. 

A wave travelling opposite to the direction of the white dwarf rotation, with a period similar to the spin period is the most plausible explanation. This wave could be the result of a considerable splitting in the \textit{g}-modes, caused by the rapid rotation of the white dwarf.

We identified in all STIS and COS spectra a small flux contribution of a second featureless continuum component that can be modelled by either flat F$_{\lambda}$ model or by blackbody emission, however its origin remains unclear.

\section*{ACKNOWLEDGEMENTS}

The authors gratefully acknowledge the constructive comments offered by the referee. The research leading to these results has received funding from the European Research Council under the European Union's Seventh Framework Programme (FP/2007-2013) / ERC Grant Agreement n. 320964 (WDTracer). Support for this work was provided by NASA through Hubble Fellowship grant \#HST-HF2-51357.001-A, awarded by the Space Telescope Science Institute, which is operated by the Association of Universities for Research in Astronomy, Incorporated, under NASA contract NAS5-26555. This work is based on observations made with the NASA/ESA Hubble Space Telescope, obtained at the Space Telescope Science Institute, which is operated by the Association of Universities for Research in Astronomy, Inc., under NASA contract NAS 5-26555. These observations are associated with programme \#12870. We thank Fondecyt for their support under the grants 3130559 (MZ) and 1141269 (MRS). MRS also thanks for support from Millennium Nucleus RC130007 (Chilean Ministry of Economy). O. Toloza was funded by the CONICYT becas-CONICYT/Beca-de-Doctorado-en-el-extranjero 72140362. Finally, we thanks to the \textit{AAVSO} observers for their contribution in this research.







\bibliographystyle{mnras}	
\bibliography{aabib_new}		

\begin{thebibliography}{}
\makeatletter
\relax
\def\mn@urlcharsother{\let\do\@makeother \do\$\do\&\do\#\do\^\do\_\do\%\do\~}
\def\mn@doi{\begingroup\mn@urlcharsother \@ifnextchar [ {\mn@doi@}
  {\mn@doi@[]}}
\def\mn@doi@[#1]#2{\def\@tempa{#1}\ifx\@tempa\@empty \href
  {http://dx.doi.org/#2} {doi:#2}\else \href {http://dx.doi.org/#2} {#1}\fi
  \endgroup}
\def\mn@eprint#1#2{\mn@eprint@#1:#2::\@nil}
\def\mn@eprint@arXiv#1{\href {http://arxiv.org/abs/#1} {{\tt arXiv:#1}}}
\def\mn@eprint@dblp#1{\href {http://dblp.uni-trier.de/rec/bibtex/#1.xml}
  {dblp:#1}}
\def\mn@eprint@#1:#2:#3:#4\@nil{\def\@tempa {#1}\def\@tempb {#2}\def\@tempc
  {#3}\ifx \@tempc \@empty \let \@tempc \@tempb \let \@tempb \@tempa \fi \ifx
  \@tempb \@empty \def\@tempb {arXiv}\fi \@ifundefined
  {mn@eprint@\@tempb}{\@tempb:\@tempc}{\expandafter \expandafter \csname
  mn@eprint@\@tempb\endcsname \expandafter{\@tempc}}}

\bibitem[\protect\citeauthoryear{{Araujo-Betancor}, {G{\"a}nsicke}, {Long},
  {Beuermann}, {de Martino}, {Sion}  \& {Szkody}}{{Araujo-Betancor}
  et~al.}{2005}]{araujo-betancoretal05-2}
{Araujo-Betancor} S.,  {G{\"a}nsicke} B.~T.,  {Long} K.~S.,  {Beuermann} K.,
  {de Martino} D.,  {Sion} E.~M.,   {Szkody} P.,  2005, \mn@doi [\apj]
  {10.1086/427914}, \href {http://adsabs.harvard.edu/abs/2005ApJ...622..589A}
  {622, 589}

\bibitem[\protect\citeauthoryear{{Arras}, {Townsley}  \& {Bildsten}}{{Arras}
  et~al.}{2006}]{arrasetal06-1}
{Arras} P.,  {Townsley} D.~M.,   {Bildsten} L.,  2006, \mn@doi [\apjl]
  {10.1086/505178}, \href {http://adsabs.harvard.edu/abs/2006ApJ...643L.119A}
  {643, L119}

\bibitem[\protect\citeauthoryear{{Bell}, {Hermes}, {Bischoff-Kim}, {Moorhead},
  {Montgomery}, {{\O}stensen}, {Castanheira}  \& {Winget}}{{Bell}
  et~al.}{2015}]{Belletal-2015}
{Bell} K.~J.,  {Hermes} J.~J.,  {Bischoff-Kim} A.,  {Moorhead} S.,
  {Montgomery} M.~H.,  {{\O}stensen} R.,  {Castanheira} B.~G.,   {Winget}
  D.~E.,  2015, \mn@doi [\apj] {10.1088/0004-637X/809/1/14}, \href
  {http://adsabs.harvard.edu/abs/2015ApJ...809...14B} {809, 14}

\bibitem[\protect\citeauthoryear{{Bergeron} et~al.,}{{Bergeron}
  et~al.}{2011}]{bergeronetal2011}
{Bergeron} P.,  et~al., 2011, \mn@doi [\apj] {10.1088/0004-637X/737/1/28},
  \href {http://adsabs.harvard.edu/abs/2011ApJ...737...28B} {737, 28}

\bibitem[\protect\citeauthoryear{{Bildsten}, {Ushomirsky}  \&
  {Cutler}}{{Bildsten} et~al.}{1996}]{bildstenetal1996}
{Bildsten} L.,  {Ushomirsky} G.,   {Cutler} C.,  1996, \mn@doi [\apj]
  {10.1086/177012}, \href {http://adsabs.harvard.edu/abs/1996ApJ...460..827B}
  {460, 827}

\bibitem[\protect\citeauthoryear{{Brickhill}}{{Brickhill}}{1983}]{brickhill-83}
{Brickhill} A.~J.,  1983, \mnras, \href
  {http://adsabs.harvard.edu/abs/1983MNRAS.204..537B} {204, 537}

\bibitem[\protect\citeauthoryear{{Brickhill}}{{Brickhill}}{1991}]{Brickhill91-2}
{Brickhill} A.~J.,  1991, \mnras, \href
  {http://adsabs.harvard.edu/abs/1991MNRAS.252..334B} {252, 334}

\bibitem[\protect\citeauthoryear{{Bullock} et~al.,}{{Bullock}
  et~al.}{2011}]{Bullocketal11-1}
{Bullock} E.,  et~al., 2011, \mn@doi [\aj] {10.1088/0004-6256/141/3/84}, \href
  {http://adsabs.harvard.edu/abs/2011AJ....141...84B} {141, 84}

\bibitem[\protect\citeauthoryear{{Chote} \& {Sullivan}}{{Chote} \&
  {Sullivan}}{2016}]{choteetal-2016}
{Chote} P.,  {Sullivan} D.~J.,  2016, \mn@doi [\mnras] {10.1093/mnras/stw421},
  \href {http://adsabs.harvard.edu/abs/2016MNRAS.458.1393C} {458, 1393}

\bibitem[\protect\citeauthoryear{{Clemens}, {van Kerkwijk}  \& {Wu}}{{Clemens}
  et~al.}{2000}]{clemensetal-2000}
{Clemens} J.~C.,  {van Kerkwijk} M.~H.,   {Wu} Y.,  2000, \mn@doi [\mnras]
  {10.1046/j.1365-8711.2000.03307.x}, \href
  {http://adsabs.harvard.edu/abs/2000MNRAS.314..220C} {314, 220}

\bibitem[\protect\citeauthoryear{{Copperwheat} et~al.,}{{Copperwheat}
  et~al.}{2009}]{copperwheatetal09-1}
{Copperwheat} C.~M.,  et~al., 2009, \mn@doi [\mnras]
  {10.1111/j.1365-2966.2008.14163.x}, \href
  {http://adsabs.harvard.edu/abs/2009MNRAS.393..157C} {393, 157}

\bibitem[\protect\citeauthoryear{{Dziembowski} \& {Koester}}{{Dziembowski} \&
  {Koester}}{1981}]{Dziembowski1981-1}
{Dziembowski} W.,  {Koester} D.,  1981, \aap, \href
  {http://adsabs.harvard.edu/abs/1981A%26A....97...16D} {97, 16}

\bibitem[\protect\citeauthoryear{{Fontaine}, {Lacombe}, {McGraw}, {Dearborn}
  \& {Gustafson}}{{Fontaine} et~al.}{1982}]{Fontaineetal1982}
{Fontaine} G.,  {Lacombe} P.,  {McGraw} J.~T.,  {Dearborn} D.~S.~P.,
  {Gustafson} J.,  1982, \mn@doi [\apj] {10.1086/160115}, \href
  {http://adsabs.harvard.edu/abs/1982ApJ...258..651F} {258, 651}

\bibitem[\protect\citeauthoryear{{Foreman-Mackey}, {Hogg}, {Lang}  \&
  {Goodman}}{{Foreman-Mackey} et~al.}{2013}]{emcee}
{Foreman-Mackey} D.,  {Hogg} D.~W.,  {Lang} D.,   {Goodman} J.,  2013, \mn@doi
  [\pasp] {10.1086/670067}, \href
  {http://adsabs.harvard.edu/abs/2013PASP..125..306F} {125, 306}

\bibitem[\protect\citeauthoryear{{G{\"a}nsicke}, {Beuermann}  \& {de
  Martino}}{{G{\"a}nsicke} et~al.}{1995}]{gaensickeetal95-1}
{G{\"a}nsicke} B.~T.,  {Beuermann} K.,   {de Martino} D.,  1995, \aap, \href
  {http://adsabs.harvard.edu/abs/1995A%26A...303..127G} {303, 127}

\bibitem[\protect\citeauthoryear{{G{\"a}nsicke}, {Hoard}, {Beuermann}, {Sion}
  \& {Szkody}}{{G{\"a}nsicke} et~al.}{1998}]{gaensickeetal98-2}
{G{\"a}nsicke} B.~T.,  {Hoard} D.~W.,  {Beuermann} K.,  {Sion} E.~M.,
  {Szkody} P.,  1998, \aap, \href
  {http://adsabs.harvard.edu/abs/1998A%26A...338..933G} {338, 933}

\bibitem[\protect\citeauthoryear{{G{\"a}nsicke}, {Szkody}, {Howell}  \&
  {Sion}}{{G{\"a}nsicke} et~al.}{2005}]{gaensickeetal05-2}
{G{\"a}nsicke} B.~T.,  {Szkody} P.,  {Howell} S.~B.,   {Sion} E.~M.,  2005,
  \mn@doi [ApJ] {10.1086/431271}, \href {2005ApJ...629..451G} {629, 451}

\bibitem[\protect\citeauthoryear{{G{\"a}nsicke}, {Long}, {Barstow}  \&
  {Hubeny}}{{G{\"a}nsicke} et~al.}{2006}]{gaensickeetal06-2}
{G{\"a}nsicke} B.~T.,  {Long} K.~S.,  {Barstow} M.~A.,   {Hubeny} I.,  2006,
  \mn@doi [\apj] {10.1086/499358}, \href
  {http://adsabs.harvard.edu/abs/2006ApJ...639.1039G} {639, 1039}

\bibitem[\protect\citeauthoryear{{Gianninas}, {Bergeron}  \&
  {Ruiz}}{{Gianninas} et~al.}{2011}]{gianninasetal11-1}
{Gianninas} A.,  {Bergeron} P.,   {Ruiz} M.~T.,  2011, \mn@doi [\apj]
  {10.1088/0004-637X/743/2/138}, \href
  {http://adsabs.harvard.edu/abs/2011ApJ...743..138G} {743, 138}

\bibitem[\protect\citeauthoryear{{Godon} \& {Sion}}{{Godon} \&
  {Sion}}{2003}]{Godonetal2003-1}
{Godon} P.,  {Sion} E.~M.,  2003, \mn@doi [\apj] {10.1086/367692}, \href
  {http://adsabs.harvard.edu/abs/2003ApJ...586..427G} {586, 427}

\bibitem[\protect\citeauthoryear{{Godon}, {Regev}  \& {Shaviv}}{{Godon}
  et~al.}{1995}]{godonetal95}
{Godon} P.,  {Regev} O.,   {Shaviv} G.,  1995, \mnras, \href
  {http://adsabs.harvard.edu/abs/1995MNRAS.275.1093G} {275, 1093}

\bibitem[\protect\citeauthoryear{{Godon}, {Sion}, {Cheng}, {Szkody}, {Long}  \&
  {Froning}}{{Godon} et~al.}{2004}]{godonetal2004-2}
{Godon} P.,  {Sion} E.~M.,  {Cheng} F.~H.,  {Szkody} P.,  {Long} K.~S.,
  {Froning} C.~S.,  2004, \mn@doi [\apj] {10.1086/422418}, \href
  {http://adsabs.harvard.edu/abs/2004ApJ...612..429G} {612, 429}

\bibitem[\protect\citeauthoryear{{Godon}, {Sion}, {Barrett}, {Hubeny},
  {Linnell}  \& {Szkody}}{{Godon} et~al.}{2008}]{godon2008A-1}
{Godon} P.,  {Sion} E.~M.,  {Barrett} P.~E.,  {Hubeny} I.,  {Linnell} A.~P.,
  {Szkody} P.,  2008, \mn@doi [\apj] {10.1086/587504}, \href
  {http://adsabs.harvard.edu/abs/2008ApJ...679.1447G} {679, 1447}

\bibitem[\protect\citeauthoryear{{Goliasch} \& {Nelson}}{{Goliasch} \&
  {Nelson}}{2015}]{goliasch+nelson15}
{Goliasch} J.,  {Nelson} L.,  2015, \mn@doi [\apj]
  {10.1088/0004-637X/809/1/80}, \href
  {http://adsabs.harvard.edu/abs/2015ApJ...809...80G} {809, 80}

\bibitem[\protect\citeauthoryear{{Hermes} et~al.,}{{Hermes}
  et~al.}{2015}]{Hermesetal-2015}
{Hermes} J.~J.,  et~al., 2015, \mn@doi [\apjl] {10.1088/2041-8205/810/1/L5},
  \href {http://adsabs.harvard.edu/abs/2015ApJ...810L...5H} {810, L5}

\bibitem[\protect\citeauthoryear{{Hilton}, {Szkody}, {Mukadam}, {Mukai},
  {Hellier}, {van Zyl}  \& {Homer}}{{Hilton} et~al.}{2007}]{Hilton2007}
{Hilton} E.~J.,  {Szkody} P.,  {Mukadam} A.,  {Mukai} K.,  {Hellier} C.,  {van
  Zyl} L.,   {Homer} L.,  2007, \mn@doi [\aj] {10.1086/521343}, \href
  {http://adsabs.harvard.edu/abs/2007AJ....134.1503H} {134, 1503}

\bibitem[\protect\citeauthoryear{{Holberg} \& {Bergeron}}{{Holberg} \&
  {Bergeron}}{2006}]{holberg+bergeron06-1}
{Holberg} J.~B.,  {Bergeron} P.,  2006, \mn@doi [\aj] {10.1086/505938}, \href
  {http://adsabs.harvard.edu/abs/2006AJ....132.1221H} {132, 1221}

\bibitem[\protect\citeauthoryear{{Hubeny} \& {Lanz}}{{Hubeny} \&
  {Lanz}}{1995}]{hubeny+lanz95-1}
{Hubeny} I.,  {Lanz} T.,  1995, \mn@doi [\apj] {10.1086/175226}, \href
  {http://adsabs.harvard.edu/abs/1995ApJ...439..875H} {439, 875}

\bibitem[\protect\citeauthoryear{{Kato}, {Maehara}  \& {Monard}}{{Kato}
  et~al.}{2008}]{katoetal2008}
{Kato} T.,  {Maehara} H.,   {Monard} B.,  2008, \mn@doi [\pasj]
  {10.1093/pasj/60.4.L23}, \href
  {http://adsabs.harvard.edu/abs/2008PASJ...60L..23K} {60, L23}

\bibitem[\protect\citeauthoryear{{Kawaler}}{{Kawaler}}{2004}]{kawaler2004}
{Kawaler} S.~D.,  2004, in {Maeder} A.,  {Eenens} P.,  eds,  IAU Symposium Vol.
  215, Stellar Rotation. p.~561

\bibitem[\protect\citeauthoryear{{Kippenhahn} \& {Thomas}}{{Kippenhahn} \&
  {Thomas}}{1978}]{kippenhahn+thomas78-1}
{Kippenhahn} R.,  {Thomas} H.-C.,  1978, \aap, \href
  {http://adsabs.harvard.edu/abs/1978A%26A....63..265K} {63, 265}

\bibitem[\protect\citeauthoryear{{Kowalski} \& {Saumon}}{{Kowalski} \&
  {Saumon}}{2006}]{kowalski+saumon2006}
{Kowalski} P.~M.,  {Saumon} D.,  2006, \mn@doi [\apjl] {10.1086/509723}, \href
  {http://adsabs.harvard.edu/abs/2006ApJ...651L.137K} {651, L137}

\bibitem[\protect\citeauthoryear{{Lenz} \& {Breger}}{{Lenz} \&
  {Breger}}{2005}]{period04}
{Lenz} P.,  {Breger} M.,  2005, \mn@doi [Communications in Asteroseismology]
  {10.1553/cia146s53}, \href
  {http://adsabs.harvard.edu/abs/2005CoAst.146...53L} {146, 53}

\bibitem[\protect\citeauthoryear{{Long}, {G{\"a}nsicke}, {Knigge}, {Froning}
  \& {Monard}}{{Long} et~al.}{2009}]{longetal2009-1}
{Long} K.~S.,  {G{\"a}nsicke} B.~T.,  {Knigge} C.,  {Froning} C.~S.,   {Monard}
  B.,  2009, \mn@doi [\apj] {10.1088/0004-637X/697/2/1512}, \href
  {http://adsabs.harvard.edu/abs/2009ApJ...697.1512L} {697, 1512}

\bibitem[\protect\citeauthoryear{{Meyer} \& {Meyer-Hofmeister}}{{Meyer} \&
  {Meyer-Hofmeister}}{1981}]{meyer+meyer-hofmeister81-1}
{Meyer} F.,  {Meyer-Hofmeister} E.,  1981, \aap, \href
  {http://adsabs.harvard.edu/abs/1981A%26A...104L..10M} {104, L10}

\bibitem[\protect\citeauthoryear{{Mukadam}, {Montgomery}, {Winget}, {Kepler}
  \& {Clemens}}{{Mukadam} et~al.}{2006}]{Mukadametal06-1}
{Mukadam} A.~S.,  {Montgomery} M.~H.,  {Winget} D.~E.,  {Kepler} S.~O.,
  {Clemens} J.~C.,  2006, \mn@doi [\apj] {10.1086/500289}, \href
  {http://adsabs.harvard.edu/abs/2006ApJ...640..956M} {640, 956}

\bibitem[\protect\citeauthoryear{{Neustroev}, {Tovmassian}, {Zharikov}  \&
  {Sjoberg}}{{Neustroev} et~al.}{2013}]{Neustroevetal2013-1}
{Neustroev} V.~V.,  {Tovmassian} G.~H.,  {Zharikov} S.~V.,   {Sjoberg} G.,
  2013, \mn@doi [\mnras] {10.1093/mnras/stt622}, \href
  {http://adsabs.harvard.edu/abs/2013MNRAS.432.2596N} {432, 2596}

\bibitem[\protect\citeauthoryear{{Piro}, {Arras}  \& {Bildsten}}{{Piro}
  et~al.}{2005}]{piroetal05-1}
{Piro} A.~L.,  {Arras} P.,   {Bildsten} L.,  2005, \mn@doi [\apj]
  {10.1086/430588}, \href {http://adsabs.harvard.edu/abs/2005ApJ...628..401P}
  {628, 401}

\bibitem[\protect\citeauthoryear{{Robinson} et~al.,}{{Robinson}
  et~al.}{1995}]{Robinsonetal95-2}
{Robinson} E.~L.,  et~al., 1995, \mn@doi [\apj] {10.1086/175132}, \href
  {http://adsabs.harvard.edu/abs/1995ApJ...438..908R} {438, 908}

\bibitem[\protect\citeauthoryear{{Schwieterman} et~al.,}{{Schwieterman}
  et~al.}{2010}]{schwieterman2010}
{Schwieterman} E.~W.,  et~al., 2010, Journal of the Southeastern Association
  for Research in Astronomy, \href
  {http://adsabs.harvard.edu/abs/2010JSARA...3....6S} {3, 6}

\bibitem[\protect\citeauthoryear{{Schwope}, {Hambaryan}, {Schwarz}, {Kanbach}
  \& {G{\"a}nsicke}}{{Schwope} et~al.}{2002}]{schwopeetal02-1}
{Schwope} A.~D.,  {Hambaryan} V.,  {Schwarz} R.,  {Kanbach} G.,
  {G{\"a}nsicke} B.~T.,  2002, \mn@doi [\aap] {10.1051/0004-6361:20011651},
  \href {http://adsabs.harvard.edu/abs/2002A%26A...392..541S} {392, 541}

\bibitem[\protect\citeauthoryear{{Sion}, {Long}, {Szkody}  \& {Huang}}{{Sion}
  et~al.}{1994}]{sionetal94-1}
{Sion} E.~M.,  {Long} K.~S.,  {Szkody} P.,   {Huang} M.,  1994, ApJ, \href
  {1994ApJ...430L..53S} {430, L53}

\bibitem[\protect\citeauthoryear{{Sion}, {Szkody}, {Cheng}, {G{\"a}nsicke}  \&
  {Howell}}{{Sion} et~al.}{2003}]{sionetal03-1}
{Sion} E.~M.,  {Szkody} P.,  {Cheng} F.,  {G{\"a}nsicke} B.~T.,   {Howell}
  S.~B.,  2003, \mn@doi [\apj] {10.1086/345445}, \href
  {http://adsabs.harvard.edu/abs/2003ApJ...583..907S} {583, 907}

\bibitem[\protect\citeauthoryear{{Szkody} \& {Mattei}}{{Szkody} \&
  {Mattei}}{1984}]{szkody+mattei84-1}
{Szkody} P.,  {Mattei} J.~A.,  1984, PASP, \href {1984PASP...96..988S} {96,
  988}

\bibitem[\protect\citeauthoryear{{Szkody}, {G{\"a}nsicke}, {Howell}  \&
  {Sion}}{{Szkody} et~al.}{2002}]{szkodyetal02-4}
{Szkody} P.,  {G{\"a}nsicke} B.~T.,  {Howell} S.~B.,   {Sion} E.~M.,  2002,
  \mn@doi [\apjl] {10.1086/342916}, \href
  {http://adsabs.harvard.edu/abs/2002ApJ...575L..79S} {575, L79}

\bibitem[\protect\citeauthoryear{{Szkody} et~al.,}{{Szkody}
  et~al.}{2010a}]{Szkodyetal10-2}
{Szkody} P.,  et~al., 2010a, \mn@doi [\apj] {10.1088/0004-637X/710/1/64}, \href
  {http://adsabs.harvard.edu/abs/2010ApJ...710...64S} {710, 64}

\bibitem[\protect\citeauthoryear{{Szkody} et~al.,}{{Szkody}
  et~al.}{2010b}]{szkodyetal10-1}
{Szkody} P.,  et~al., 2010b, \mn@doi [\apj] {10.1088/0004-637X/716/2/1531},
  \href {http://adsabs.harvard.edu/abs/2010ApJ...716.1531S} {716, 1531}

\bibitem[\protect\citeauthoryear{{Szkody} et~al.,}{{Szkody}
  et~al.}{2012}]{szkodyetal12-1}
{Szkody} P.,  et~al., 2012, \mn@doi [\apj] {10.1088/0004-637X/753/2/158}, \href
  {http://adsabs.harvard.edu/abs/2012ApJ...753..158S} {753, 158}

\bibitem[\protect\citeauthoryear{{Templeton}, {Stubbings}, {Waagen}, {Schmeer},
  {Pearce}  \& {Nelson}}{{Templeton} et~al.}{2007}]{Templetonetal2007-cbet}
{Templeton} M.,  {Stubbings} R.,  {Waagen} E.~O.,  {Schmeer} P.,  {Pearce} A.,
   {Nelson} P.,  2007, Central Bureau Electronic Telegrams, \href
  {http://adsabs.harvard.edu/abs/2007CBET..922....1T} {922, 1}

\bibitem[\protect\citeauthoryear{{Thorstensen}}{{Thorstensen}}{2003}]{Thorstensen03-3}
{Thorstensen} J.~R.,  2003, \mn@doi [\aj] {10.1086/379308}, \href
  {http://adsabs.harvard.edu/abs/2003AJ....126.3017T} {126, 3017}

\bibitem[\protect\citeauthoryear{{Thorstensen}, {Patterson}, {Kemp}  \&
  {Vennes}}{{Thorstensen} et~al.}{2002}]{thorstensenetal02-2}
{Thorstensen} J.~R.,  {Patterson} J.,  {Kemp} J.,   {Vennes} S.,  2002, \mn@doi
  [\pasp] {10.1086/342484}, \href
  {http://adsabs.harvard.edu/abs/2002PASP..114.1108T} {114, 1108}

\bibitem[\protect\citeauthoryear{{Tovmassian} et~al.,}{{Tovmassian}
  et~al.}{2003}]{tovmassianetal03-1}
{Tovmassian} G.,  et~al., 2003, \mn@doi [\pasp] {10.1086/375031}, \href
  {http://adsabs.harvard.edu/abs/2003PASP..115..725T} {115, 725}

\bibitem[\protect\citeauthoryear{{Townsley} \& {Bildsten}}{{Townsley} \&
  {Bildsten}}{2004}]{townsley+bildsten-2004}
{Townsley} D.~M.,  {Bildsten} L.,  2004, \mn@doi [\apj] {10.1086/379647}, \href
  {http://adsabs.harvard.edu/abs/2004ApJ...600..390T} {600, 390}

\bibitem[\protect\citeauthoryear{{Townsley} \& {G{\"a}nsicke}}{{Townsley} \&
  {G{\"a}nsicke}}{2009}]{townsley+gaensicke09-1}
{Townsley} D.~M.,  {G{\"a}nsicke} B.~T.,  2009, \mn@doi [\apj]
  {10.1088/0004-637X/693/1/1007}, \href
  {http://adsabs.harvard.edu/abs/2009ApJ...693.1007T} {693, 1007}

\bibitem[\protect\citeauthoryear{{Townsley}, {Arras}  \& {Bildsten}}{{Townsley}
  et~al.}{2016}]{townsleyetal-16}
{Townsley} D.~M.,  {Arras} P.,   {Bildsten} L.,  2016, preprint, \href
  {http://adsabs.harvard.edu/abs/2016arXiv160102046T} {} (\mn@eprint {arXiv}
  {1601.02046})

\bibitem[\protect\citeauthoryear{{Tremblay}, {Ludwig}, {Steffen}, {Bergeron}
  \& {Freytag}}{{Tremblay} et~al.}{2011}]{Tremblayetal11-1}
{Tremblay} P.-E.,  {Ludwig} H.-G.,  {Steffen} M.,  {Bergeron} P.,   {Freytag}
  B.,  2011, \mn@doi [\aap] {10.1051/0004-6361/201117310}, \href
  {http://adsabs.harvard.edu/abs/2011A%26A...531L..19T} {531, L19}

\bibitem[\protect\citeauthoryear{{Unno}, {Osaki}, {Ando}, {Saio}  \&
  {Shibahashi}}{{Unno} et~al.}{1989}]{Unnoetal-1989}
{Unno} W.,  {Osaki} Y.,  {Ando} H.,  {Saio} H.,   {Shibahashi} H.,  1989,
  {Nonradial oscillations of stars}

\bibitem[\protect\citeauthoryear{{Uzdensky}}{{Uzdensky}}{2002}]{Uzdensky2002-1}
{Uzdensky} D.~A.,  2002, \mn@doi [\apj] {10.1086/340308}, \href
  {http://adsabs.harvard.edu/abs/2002ApJ...572..432U} {572, 432}

\bibitem[\protect\citeauthoryear{{Van Grootel}, {Fontaine}, {Brassard}  \&
  {Dupret}}{{Van Grootel} et~al.}{2015}]{VanGrooteletal2015}
{Van Grootel} V.,  {Fontaine} G.,  {Brassard} P.,   {Dupret} M.-A.,  2015,
  \mn@doi [\aap] {10.1051/0004-6361/201425386}, \href
  {http://adsabs.harvard.edu/abs/2015A%26A...575A.125V} {575, A125}

\bibitem[\protect\citeauthoryear{{Vican} et~al.,}{{Vican}
  et~al.}{2011}]{Vicanetal11-1}
{Vican} L.,  et~al., 2011, \mn@doi [\pasp] {10.1086/662633}, \href
  {http://adsabs.harvard.edu/abs/2011PASP..123.1156V} {123, 1156}

\bibitem[\protect\citeauthoryear{{Warner}}{{Warner}}{1995}]{warner95-1}
{Warner} B.,  1995, Cataclysmic Variable Stars.
Cambridge University Press, Cambridge

\bibitem[\protect\citeauthoryear{{Woudt} \& {Warner}}{{Woudt} \&
  {Warner}}{2002}]{woudt+warner02-2}
{Woudt} P.~A.,  {Warner} B.,  2002, \mn@doi [\apss] {10.1023/A:1020809407912},
  \href {http://adsabs.harvard.edu/abs/2002Ap%26SS.282..433W} {282, 433}

\bibitem[\protect\citeauthoryear{{Wu} \& {Goldreich}}{{Wu} \&
  {Goldreich}}{1999}]{wu+goldreich-99}
{Wu} Y.,  {Goldreich} P.,  1999, \mn@doi [\apj] {10.1086/307412}, \href
  {http://adsabs.harvard.edu/abs/1999ApJ...519..783W} {519, 783}

\bibitem[\protect\citeauthoryear{{van Spaandonk}, {Steeghs}, {Marsh}  \&
  {Parsons}}{{van Spaandonk} et~al.}{2010}]{vanSpaandonketal2010-2}
{van Spaandonk} L.,  {Steeghs} D.,  {Marsh} T.~R.,   {Parsons} S.~G.,  2010,
  \mn@doi [\apjl] {10.1088/2041-8205/715/2/L109}, \href
  {http://adsabs.harvard.edu/abs/2010ApJ...715L.109V} {715, L109}

\bibitem[\protect\citeauthoryear{{van Zyl}, {Warner}, {O'Donoghue}, {Sullivan},
  {Pritchard}  \& {Kemp}}{{van Zyl} et~al.}{2000}]{vanzyletal00-1}
{van Zyl} L.,  {Warner} B.,  {O'Donoghue} D.,  {Sullivan} D.,  {Pritchard} J.,
   {Kemp} J.,  2000, Baltic Astronomy, \href
  {http://adsabs.harvard.edu/abs/2000BaltA...9..231V} {9, 231}

\bibitem[\protect\citeauthoryear{{van Zyl} et~al.,}{{van Zyl}
  et~al.}{2004}]{vanzyletal04-1}
{van Zyl} L.,  et~al., 2004, \mn@doi [\mnras]
  {10.1111/j.1365-2966.2004.07646.x}, \href
  {http://adsabs.harvard.edu/abs/2004MNRAS.350..307V} {350, 307}

\makeatother
\end{thebibliography}



\appendix




\bsp	
\label{lastpage}
\end{document}